# The Universal Arrow of Time V-VI: (Part V) Unpredictable dynamics (Part VI) Future of artificial intelligence - Art, not Science: Practical Application of Unpredictable Systems

## The Universal Arrow of Time V: Unpredictable dynamics.

Kupervasser Oleg

## Abstract

We see that the exact equations of quantum and classical mechanics describe ideal dynamics which is reversible and result in Poincare's to returns. Real equations of physics describe observable dynamics. It is, for example, master equations of the statistical mechanics, hydrodynamic equations of viscous fluid, Boltzmann equation in thermodynamics, and the entropy increase law in the isolated systems. These laws are nonreversible and exclude Poincare's returns to an initial state. Besides these equations describe systems in terms of macroparameters or phase distribution functions of microparameters. Two reasons of such differences between ideal and observable dynamicses exist. At first, it is uncontrollable noise from the external observer. Secondly, when the observer is included into described system (introspection) the complete self-description of a state of such full system is impossible. Besides introspection is possible during finite time when the thermodynamic time arrow of the observer exists and does not change the direction. Not for all cases broken by external noise (or incomplete at introspection) ideal dynamics can be changed to predictable observable dynamics. For many systems introduction of the macroparameters that allow exhaustively describe dynamics of the system, is impossible. Their dynamics to become in principle unpredictable, sometimes even unpredictable by the probabilistic way. We will name dynamics describing such system, *unpredictable dynamics*. As follows from the definition of such systems, it is impossible to introduce a complete set of macroparameters for *unpredictable dynamics*. (Such set of macroparameters for observable dynamics allowed to predict their behavior by a complete way.) Dynamics of unpredictable systems is not described and not predicted *by scientific* methods. Thus, **the science itself puts boundaries for its applicability.** But such systems can *intuitively* «to understand itself» and «to predict» the own behavior or even «to communicate with each other» at *intuitive* level.

## 1. Introduction

Let's give definitions *observed and ideal dynamicses* [1-4], and also we will explain necessity of introduction of observable dynamics. We will name as ideal dynamics exact laws of quantum or classical mechanics. Why we have named their ideal? Because for the most of real systems the entropy increase law or wave packet reduction in the quantum case. These properties contradict with laws of ideal dynamics. Ideal dynamics is reversible and includes Poincare's returns. It is not observed in nonreversible observable dynamics. Whence there is this inconsistency between the dynamicses?

The real observer is always macroscopic system far from thermodynamic equilibrium. It possesses a thermodynamic time arrow which exists finite time (before the equilibrium reaching) and can change its direction. Besides, there is a small interaction of the observer with observable system which results in alignment of thermodynamic time arrows and, in case of a quantum mechanics, in wave packet reduction.

The observer describes the observable system in terms of macroparameters and corespondent thermodynamic time arrow. It also results in the difference of observable dynamics and ideal dynamics. The ideal dynamics is formulated with respect to the abstract coordinate time in terms of microparameters.

Violations of ideal dynamics are related to either openness of measured systems (i.e. can be explained by influence of environment/observer) or impossibility of self-measuring at introspection (for the full closed physical systems including both the environment and the observer). What is possible to do for such cases? The real system is either open or incomplete, i.e. we can not use physics for perdition of the system evolution? Not so!

Lots of such systems can be described by equations of exact or probabilistic dynamics, in spite of openness or description incompleteness. We name it observable dynamics. The most of equations in physics - master equations of statistical mechanics, hydrodynamic equation of viscous fluid, Boltzmann equation in thermodynamics, and the entropy increase law are equations of observable dynamics.

To possess the property specified above observable dynamics should answer certain requirements. It cannot operate with the full set of microvariables. In observable dynamics we use much smaller number of macrovariables which are some functions of microvariables. It makes the dynamics much more stable with respect to errors of initial conditions and external noise. Really, a microstate change does not result inevitably in a macrostate change. Since one macrostate is correspondent to a huge set of microstates. For gas macrovariables are, for example, the density, pressure, temperature and entropy. Microvariables are velocities and coordinates of all its molecules.

How to get observable dynamics from ideal dynamics? It can be gotten either by insertion to equations of the ideal equations small external noise, or insertion of errors to an initial state. Errors/noise should be large enough to break effects unobservable in reality. It is reversibility of motion or Poincare's returns. On the other hand they should be small enough not to influence observable processes with entropy increase.

For the complete physical system including the observer, observable system and a surrounding medium Observable Dynamics is not falsifiable in Popper's sense [36] (under condition of fidelity of Ideal Dynamics). I.e. the difference between Ideal and Observable Dynamics in this case cannot be observed in experiment.

However, there exist cases when it is not possible to find any observable dynamics. The system are unpredictable, because of either openness or description incompleteness. It is a case of *unpredictable dynamics* [21, 29-33], considered here.

## 2. Unpredictable dynamics

Let's introduce concept *synergetic models* [10]. We will name so simple physical or mathematical systems. Such systems illustrate in a simple form some real or supposed properties of unpredictable and complex (living) systems.

Unpredictable systems, as a result of its unpredictability, are extremely unstable with respect to external observation or thermal noise. To prevent their chaotization, they should have some protection from external influence.

Therefore we mainly interested in synergetic models of systems that are capable to protect itself from external noise (from decoherence in a quantum mechanics). They conserve internal correlations (quantum or classical), resulting in reversibility or Poincare's returns. Also they can conserve the correlations with surrounding world.

There are three methods for such protection:

1) The passive method - creation of some "walls" or shells impenetrable for noise. It is possible to keep also such systems at very low temperatures. An example is many models of quantum computers.

2) The active method, inverse to passive - complex dissipative or living systems, they conserve disequilibrium by the help of an active interaction and an interchanging of energy and substance with environment (metabolism). It is thought, that the future models of quantum computers should correspond to this field.

3) When correlations cover the whole Universe. The external source of noise is absent here. Origin of correlations over Universe is that Universe was in low entropy initial states. Universe appeared from Big Bang. We will name these correlations as global correlations. Sometimes it is figuratively named «holographic model of Universe»

Three facts ought to be noted:

1) Many complex systems during evolution pass dynamic bifurcation points. There are several alternative ways of future evolution. The selection one from them depends on the slightest fluctuations of the system state in the bifurcation point [5-6]. In these points even weak correlations can have huge influence on future. These correlations define one from alternative ways of future evolution specified above. Presence of such correlations restricts predictive force of the Science, but it does not restrict at all our personal intuition. Since we are an integral part of our Universe we are capable at some subjective level to "feel" these correlations inaccessible for scientific observation. No contradiction with current physics exists here.
2) In described unobservable systems the entropy decrease is often observed or they are supported at a very low-entropy state. It does not contradict to the second thermodynamics law of the entropy decrease. Really, for creation of both passive and the active protection huge negentropy from environment is necessary. Therefore the total entropy of system and an environment only increase. The entropy increase law remains correct for full system (observable system + an environment + the observer) though it is untrue for the observable system. Entropy decrease in full system can happen, for example Poincare's returns. But they are unobservable [1-4]. Therefore we can skip them.
3) Existence of many unpredictable systems is accompanied by the entropy decrease (It does not contradict to the entropy increase according to the second law of thermodynamics as it is explained above in the third item). Thus, existence of such systems correspondent to the generalized principle Le-Shatelie - Brown: the system hinders with any modification of the state caused both external action, and internal processes, or, otherwise, any modification of a state of the system, caused both external, and internal reasons, generates in system the processes guided on reducing this modification. In this case the entropy growth generates appearance of systems cause the entropy decrease.
4) Often maximum entropy production principle (MaxEPP) demonstrates correct results [38]. According to this principle the nonequilibrium system to aspire to a state at which entropy growth in system would be maximal. Despite an apparent inconsistency, MaxEPP does not contradict to Prigogine's minimum entropy production principle (MinEPP) for the linear nonequilibrium systems [38]. These are absolutely different variation principles. Though for both case the extreme of the same function (the entropy production) is looking for, but various restrictions and various parameters of a variation are thus used. It is not necessary to oppose these principles, as they are applicable to various stages of evolution of nonequilibrium system. MaxEPP means, that dissipative unpredictable systems (including living systems), being in the closed system with finite volume, accelerate appearance of thermodynamic equilibrium for this system. It means that they reduce also Poincare's return time, i.e. promote faster return to the low-entropy state. It again corresponds to the generalized principle Le-Shatelie - Brown: the entropy growth generates appearance of systems cause the entropy decrease. From all above-stated it is possible to give very interesting conclusion: *global "purpose" of dissipative systems (including living systems) is (a) minimization of their own entropy (b) stimulation of the global full system to faster Poincare's return to the initial low-entropy state.*

5) Global correlations generally "spread" over the closed system with finite volume and result only in Poincare's unobservable return [1-4]. However in the presence of objects conserving local correlations, global correlations can become apparent in correlation between such objects with each other and around world. Thus, presence of conserved local correlations allows to make global correlations to be observable, preventing their full "spreading" over the system.
6) Correct definition of thermodynamic macroscopic entropy is very difficult problem for complex physical systems without local equilibrium [39].
7) Very important facts ought to be noted. Unstable correlations exist not only in quantum, but also in the classical mechanics. Hence, such models should not have only quantum character. They can be also classical! Very often wrongly it is wrongly stated, that only the quantum mechanics have such properties **[11-12].** However**,** it is not so **[7-9].** Introduction by "hands" small, but finite interaction during classical measurement and small errors of an initial state erases the difference between properties of quantum and classical mechanics (in the presence of unstable correlations of microstates).

# 3. Synergetic models of local correlations

Let's consider examples of the synergetic models of unpredictable systems using the passive or active methods for protection from noise.

1) There are unusual cases for which there is no alignment of thermodynamic time arrows [12-13].

2) Phase transition or bifurcation points. In such point (some instance for evolution or some value for external parameter) macroscopic system, described by observable dynamics, can transform to not single, but several macroscopic states. That is, in these points observable dynamics loses the unambiguity. In these points there are huge macroscopic fluctuations, and used macroparameters does not result in predictability of the system. Evolution becomes unpredictable, i.e. there is unpredictable dynamics.

3) Let take quantum microscopic or mesoscopic system described by ideal dynamics and isolated from decohernece. Its dynamics depends on uncontrollable microscopic *quantum correlations*. These correlations are very unstable and can disappear as result of decoherence (entangling with environment/observer). For example, let us consider quantum system. Suppose that some person knows its initial and final state. Its behavior is completely predicted by such person. In a time interval between start and final the system is isolated from an environment/observer. In that case these microscopic correlations do not disappear and influence dynamics. However for the second person who is not present at start, its behavior is *uncertain* and *unpredictable*. Moreover, an attempt of the second person to observe some intermediate state of the quantum computer would result in destroying its normal operation. I.e. from the point of view of such observer there is unpredictable dynamics. Well-known examples of such systems are *quantum computers* and *quantum cryptographic transmitting systems* **[14-15].**

Quantum computer is unpredictable for any observer who does not know its state in the beginning of calculations. Any attempt of such observer to measure intermediate state of quantum computer during calculation destroys calculation process in unpredictable way. Its other important property is high parallelism of calculation. It is a consequence of QM laws linearity. Initial state can be chosen as the sum of many possible initial states of "quantum bits of the information". Because of QM laws linearity all components of this sum can evolve in independent way. This parallelism allows solving very quickly many important problems which usual computer can not solves over real time. It gives rise to large hopefulness about future practical use of quantum computers.

Quantum cryptographic transmitting systems use property of the unpredictability and unobservability of “messages” that can not be read during transmitting by any external observer. Really, these “messages” are usual quantum systems featured by quantum laws and quantum correlations. An external observer which has no information about its initial states and try make measuring (reading) of "message" over transmission inevitably destroy this transmission. Thus, message interception appears *principally* impossible under physics laws.

4) It should be emphasized, contrary to the widespread opinion, that both quantum computers and quantum cryptography **[14-15]** have classical analogues. Really, in classical systems, unlike in quantum systems measuring can be made precisely in principle without any measured state distorting. However, in classical chaotic systems as well there are the uncontrollable and unstable microscopic additional correlations resulting in reversibility and Poincare's returns. Introducing “by hands” some small finite perturbation or initial state errors destroys these correlations and erases this principal difference between classical and quantum system behavior. Such small external noise from environment always exists in any real system. By isolation of chaotic classical systems from this external noise we obtain classical analogues of isolated quantum devices with quantum correlations.

There exist synergetic models of the classical computers which ensure, like quantum computers, huge parallelism of calculations **[7]**

Analogues of quantum computers are the molecular computers **[9]**. The huge number of molecules ensures parallelism of evaluations. The unstable microscopic additional correlations (resulting in reversibility and returns) ensure dynamics of intermediate states to be unpredictable for the external observer which is not informed about the computer initial state. He would destroy computer calculation during attempt to measure some intermediate state. .

Similar arguments can be used for classical cryptographic transmitting systems using these classical unstable microscopic additional correlations for information transition. "Message" is some classical system that is chaotic in intermediate states. So any attempt to intercept it inevitably destroys it similarly to QM case.

5) Conservation of unstable microscopic correlations can be ensured not only by passive isolation from an environment and the observer but also by active dynamic mechanism of perturbations cancelling. It happens in so-called physical **stationary systems** in which steady state is supported by continuous **stream of energy or substance through system**. An example is a micromaser **[16]** - a small and well conducting cavity with electromagnetic radiation inside. The size of a cavity is so small that radiation is necessary to consider with the help of QM. Radiation damps because of interaction with conducting cavity walls. This system is well featured by density matrix in base energy eigenfunction. Such set is the best choose for observable dynamics. Microscopic correlations correspond to nondiagonal elements of the density matrix. Nondiagonal elements converge to zero much faster than diagonal ones during radiation damping. In other words, decoherence time is much less than relaxation time. However, beam of excited particles, passing through a micromaser, leads to the strong damping deceleration of density matrix nondiagonal elements (microcorrelations). It also leads to non-zero radiation in steady state.

Also in the theory of quantum computers methods of the active protection are developed. These methods protect quantum correlations from decoherence. They are capable to conserve correlations as long as desired, by iterating cycles of active quantum error correction. Repetition code in quantum information is not possible due to the no-cloning theorem. Peter Shor first discovered method of formulating a quantum error correcting code by storing the information of one qubit onto a highly-entangled state of nine qubits [17].

6) In physics usually a macrostate is considered as some passive function of a microstate. However it is possible to consider a case when the system observes (measures) both its macrostate and an environment macrostate. The result of the observation (measurement) is recorded into microscopic "memory". By such a way the feedback appears between macrostates and microstates.

An example of very complex stationary systems is living systems. Their states are very far from thermodynamic equilibrium and extremely complex. These systems are high ordered but their order is strongly different from an order of lifeless periodical crystal. Low entropy disequilibrium of the live is supported by entropy growth in environment[1]. It is metabolism - the continuous stream of substance and energy through a live organism. On the other hand, not only metabolism supports disequilibrium, this disequilibrium is himself catalytic agent of metabolic process, i.e. creates and supports it at necessary level. As the state of live systems is strongly nonequilibrium, it can support existing unstable microcorrelations, disturbing to decoherence. These correlations can be both between parts of live system, and between different live systems (or live systems with lifeless system). If it happens dynamics of live system can be referred to as unpredictable dynamics. Huge successes of the molecular biology allow describing very well dynamics of live systems. But there are no proof that we capable to feature completely all very complex processes in live system.

It is difficult enough to analyze real living systems within framework of concepts of ideal, observed and unpredictable dynamicses because of their huge complexity. But it is possible to construct simple mathematical models. It is, for example, nonequilibrium stationary systems with metabolism. It would allow us to understand a possible role of all of three dynamicses for such systems. These models can be both quantum [11-12, 18-20, 35] and classical [7-9].

7) Described above cases do not characterize all multiplicity of unpredictable dynamicses. The exact conditions at which ideal dynamics transfers in observable and unpredictable dynamics is completely not solved problem for mathematics and physics yet. Also there is not solved problem (connected to the previous problem) about a role of these of three dynamicses for complex stationary systems. The solution of these problems will allow to understand more deeply physical principles of life.

## 4. Synergetic models of global correlations expanded over the whole Universe.

With the help synergetic "toy" models it is possible to understand synchronicity[2] (simultaneity) of processes causally not connected [37], and also to illustrate a phenomenon of the global correlations.

Global correlations of the Universe and the definition of life as the totality of systems maintaining correlation in contrast to the external noise is a reasonable explanation of the mysterious silence of Cosmos, i.e. the absence of signals from other intelligent worlds. All parts of the universe, having the unique center of origin (Big Bang), are correlated, and life maintains these correlations which are at the base of its existence. Therefore the emergence of life in different parts of the Universe is correlated, so that all the civilizations have roughly the same

---

[1] *So, for example, entropy of epy Sun grows. It is an energy source for life on the Earth.*

[2] The study conducted by Russian specialists under the guidance of Valeri Isakov mathematics, which specializes in paranormal phenomena. Data from domestic flights they could not be obtained, so the researchers used Western statistics. As it turned out, over the past 20 years of flight, which ended in disaster, refused on 18% more people than normal flights. "We are just mathematics, which revealed a clear statistical anomaly. But mystically-minded people may well associate it with the existence of some higher power "- quoted Isakov," Komsomolskaya Pravda ". http://mysouth.su/2011/06/scientists-have-proved-the-existence-of-guardian-angels/; http://kp.ru/daily/25707/908213/

“That was Staunton’s theory, and the computer bore him out. In cases where planes or trains crash, the vehicles are running at 61 percent capacity, as regards passenger loads. In cases where they don’t, the vehicles are running at 76 per cent capacity. That’s a difference of 15 percent over a large computer run, and that sort of across-the-board deviation is significant. Staunton points out that, statistically speaking, a 3 percent deviation would be food for thought, and he’s right. It’s an anomaly the size of Texas. Staunton’s deduction was that people know which planes and trains are going to crash… that they are unconsciously predicting the future."
Stephen King, "The Stand" (1990)

level of development, and there are just no supercivilizations capable of somehow reaching Earth.

## 4.1 Blow up systems

Example are nonstationary systems with "blow up" **[6, 22-25],** considered Kurdumov's school. In these processes a function on plane is defined. Its dynamics is described by the non-linear equation, similar to the burning equation:

$$\partial\rho / \partial t = f(\rho) + \partial / \partial r \, (H(\rho)\, \partial\rho / \partial r), \qquad \text{(I)}$$

where ρ - a density, $N = \int \rho \, dr$, $r$ - space coordinate, $t$ - time coordinate, $f(\rho)$, $H(\rho)$ - non-linear connections:

$f(\rho) \to \rho^{\beta}$, $H(\rho) \to \rho^{\sigma}$,

These equations have a set of the dynamic solutions, named solutions with "blow up". It was proved localization of processes in the form of structures (at $\beta > \sigma + 1$) with discrete spectrum. The structures can be simple (with individual maximums of different intensity). They also can be complex (united simple structures) with different space forms and several maximums of different intensity. It is shown, that the non-linear dissipative medium potentially contains a spectrum of such various structures-attractors. Let ($r$, $\varphi$) be polar coordinateses.

$$\rho(r,\varphi,t) = g(t)\Theta_i(\xi,\varphi), \quad \xi = \frac{r}{\psi(t)}, \quad 1 < i < N$$

$$g(t) = \left(1 - \frac{t}{\tau}\right)^{-\frac{1}{\beta-1}}, \quad \psi(t) = \left(1 - \frac{t}{\tau}\right)^{\frac{\beta-\sigma-1}{\beta-1}}$$

Number of eigenfunctions:

$$N = \frac{\beta - 1}{\beta - \sigma - 1}$$

For these solutions value of function can converge to infinity for *finite* time τ. It is interesting that function reaches infinity in all maximums in the same instant, i.e. is synchronously. In process of converging to time τ the solution "shrinks", the maximums "blow up" and moves to common centre. About the moment of 0.9τ the system becomes unstable and fluctuations of the initial condition can destroy the solution. For high correlated initial condition it is possible to reduce these fluctuations to as small as desired.

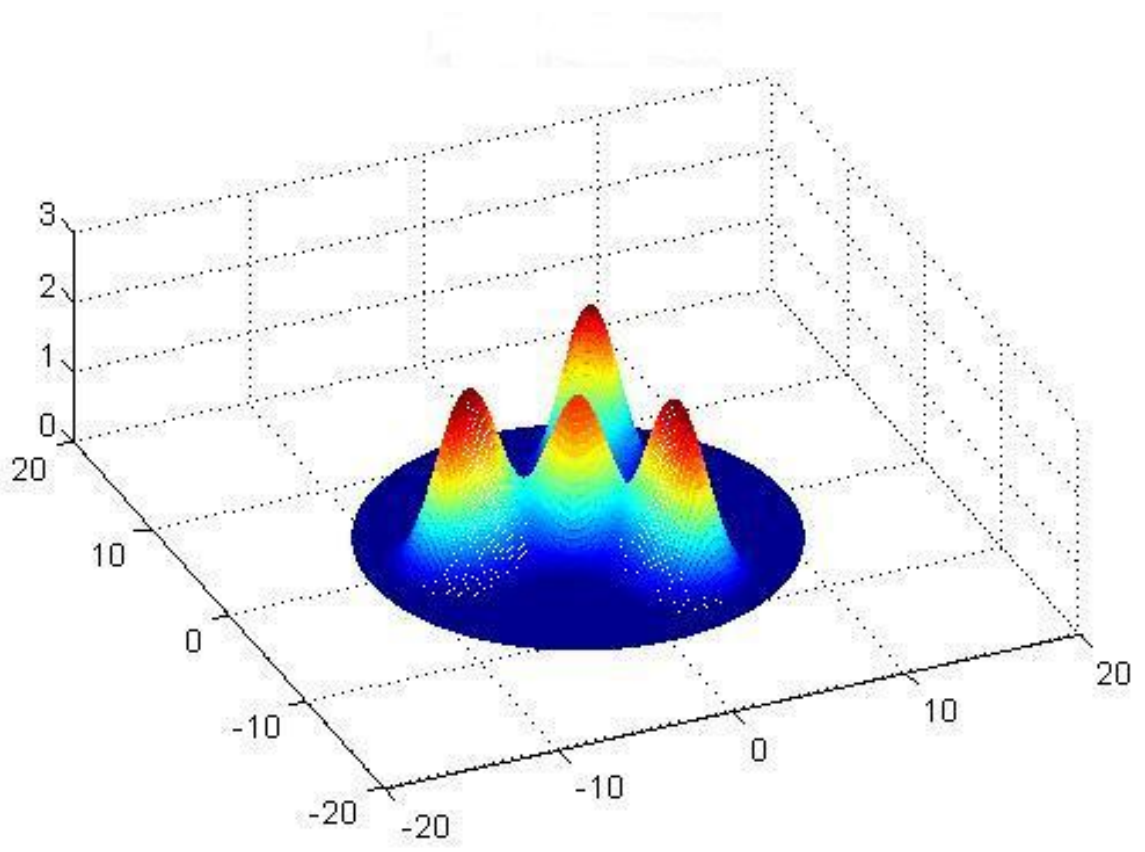

**Рис. 1** From [34]. It is one from structures-attractors of the burning equation (I) in the form of the solution with “blow up”.

By means of such models we can illustrate people population growth (or level of engineering development of civilizations) in megacities of our planet **[25].** Points of a maximum of function $\rho$ are megacities, and population density is a value of the function $\rho$.

It is possible to spread this model to the whole Universe. Then the points of a maximum are civilizations, and population density of civilizations (or level of engineering development of civilizations) is a value of the function $\rho$. For this purpose we will complicate model. Let during the moment when process starts to go out on a growing asymptotic solution there is very fast expansion ("inflation") of the plane in which process with "blow up" runs. Nevertheless, processes of converging to infinity remain synchronous and are featured by the equation of the same type (only with the changed scale) in spite of the fact that maximums are distant at large intervals.

This more difficult model is possible to explain qualitatively synchronism of processes in very far parts of our Universe as a result "inflation" after Big Bang. The high degree of global correlations reduces the fluctuations result in destroying the solution structure. These global correlations model coherence of parts of our Universe.

Processes with "blow up" appear with necessary completeness and complexity only for some narrow set of the coefficients of the equation (I). ($N \gg 1$, $\beta > \sigma + 1$, $\beta \approx \sigma + 1$ is necessary conditions for appearance of a structure with large number of maximums and their slow coming to the common center). It allows to draw an analogy with «anthropic principle» **[26].** The anthropic principle states, that the fundamental constants of the Universe have such values that a result of Universe’s evolution is our Universe with anthropic «beings», capable to observe the Universe.

It is necessary to pay attention to one fact. If we want that the ordered state in model would not be destroyed at $t=0.9\tau$, and would continue to exist as long as possible then exact adjustment is required *not only for model parameters, but also for an initial state*. It is necessary, that fluctuations arising from the initial state would not destroy orderliness as long as possible. And the presence of this rare exclusive state also can be explained by the anthropic principle.

## 4.2 «Cellular» model of Universe.

Also it is interesting to illustrate the complex processes by means of "cellular" model. Discrete Hopfield's model [27-28] can be used as a good basis. This model can be interpreted as a neural network with a feedback or as a spin lattice (a spin glass) with unequal interactions between spins. Such systems is used for a pattern recognition.

This system can be featured as a square two-dimensional lattice of meshes *NxN* which can be either black, or white ($S_i=\pm 1$). Coefficients of the linear interaction between meshes $J_{ji}$ are unequal for different pairs of meshes. They can be chosen so, that in the process of discrete evolution the overwhelming majority of initial states transfers in one of possible final states. This set of final states (attractors) can be chosen and defined "by hands".

$$S_i(t+1) = sign\left[\sum_{j=1}^{N} J_{ij} S_j(t)\right], \quad 1 \le i \le N$$

$$J_{ij} = J_{ji}, \; J_{kk} = 0 \quad 1 \le i, j, k \le N$$

Attractors correspond to energy *E* minimum:

$$E = -\frac{1}{2}\sum_{i=1}^{N}\sum_{j=1}^{N} J_{ij} S_i S_j$$

Let choose lattice attractors to be letters A or B.

There are such two initial unstable states which differ only on one mesh (*a critical element*). Thus one of them has a state as an attractor A, and another - B. Such unstable initial states well illustrates a property of the *global instability* of a complex system. This instability is inherent in a system as a whole, not in its some part. Only some external observer can change the value of the critical element and vary system evolution. Internal dynamics of the system cannot do it. *Global correlation* between meshes of an unstable initial state defines completely a final attractor (A or B) of the lattice.

It is possible to complicate model. Let each mesh in the lattice featured above is such sublattice. We will define evolution of such composite lattice going to two stages.

At the first stage large meshes do not interact. Interaction exists only into sublattices. This interaction is the same as for the one-stage model featured above. Coefficients of the linear interaction between meshes are chosen so that attractors, as well as earlier, are the letter A or B. Initial states of all sublattices can be chosen as unstable and containing the critical element. A final state A of sublattices we will perceive as a black mesh for a large lattice, and a state B of sublattices we will perceive as a white mesh.

The second stage of evolution is defined as evolution of this large lattice over the same way as in the one-stage model featured above. The initial state of the large lattice is defined by the first stage. This initial state, appearing at the first stage, is also unstable and contains the critical element. For final state of the large lattice to each black mesh we will appropriate a state A of the sublattices, to each white mesh we will appropriate a state B of the sublattices.

The initial state of the composite lattice can be chosen always so that attractor of the two-stage process will be A. For every mesh included to A the sublattice state also corresponds to A. Let's name this state of the composite lattice as «A-A». Then just such final attractor can be explained by:

a) the global correlations of the unstable initial state

b) the specific selection of all coefficients of interaction between meshes.

Let's complicate model even more. By analogy to the aforesaid, we will make this lattice not two-level, but three-level, and process instead of two-stage we will make three-stage . A final state will be «A-A-A».

Let's suppose, that before the beginning of the aforementioned three-stage process our composite lattice occupied very small field of physical space. But as a result of expansion ("inflation") it was dilated to huge size. Then the aforementioned three-stage process has begun. Thus it is possible to explain presence of the unstable correlation of the initial state of the composite lattice leading to a total state «A-A-A». Indeed, before "inflation" all meshes were closed each other. So the unstable initial correlation can be easily formed under such conditions.

This three-level composite lattice can be compared to our Universe. Its smallest sublattices «A» can be compared to «intelligent organisms». Lack of their interaction with an environment at the first stage (before forming of a final state «A») - is equivalent to the active or passive protection of internal correlations from external noise. Lattices of the second level in a state «A-A» correspond to "civilizations" organized by «intelligent organisms» («A») at the second stage. At the third stage "supercivilization" («A-A-A») is formed by "civilizations" («A-A»).

Then global correlations of the unstable initial state of the composite lattice can be analogue of the possible global correlations of the unstable initial state of our Universe existed before its inflation. Coefficients of interaction of the meshes correspond to the fundamental constants of our Universe. Initial process of the lattice expansion (before its three-stage evolutions) corresponds to Big Bang. The specific selection of interaction coefficients between the meshes leading to the asymptotic state «A-A-A» and the initial correlations can be explain by «anthropic principle». We remind here that the anthropic principle states, that the fundamental constants of the Universe have such values that a result of Universe’s evolution is our Universe with anthropic «beings», capable to observe the Universe.

# Acknowledgment

We thank Hrvoje Nikolic and Vinko Zlatic for discussions and debates which help very much during writing this paper.

# The Universal Arrow of Time VI:

# Future of artificial intelligence – Art, not Science or Practical Application of Unpredictable Systems.

## Kupervasser Oleg

Perspective of an artificial intellect (AI) future is considered. It is shown, that AI development in the future will be closer to art, than a science. Complex dissipative systems which behavior cannot be understood completely in principle will be a basis of AI. Nevertheless, this not complete understanding will not be a barrier for their practical use.

## Introduction

Now in the world the technologies relating to design of systems of artificial intellect (AI) actively develop. In this paper it would be desirable to consider not tactical, but strategic problems of this process. Now not many interesting papers on this topic are available, but they exist [1]. It is relating to a fact that most of serious experts is occupied by a solution of tactical problems and often does not think about farther prospects. However the situation at the beginning of cybernetics origin was not that. Then these problems were actively considered. Therefore we will construct our paper as a review of problems of cybernetics as they saw to participants of the symposium in 1961 [2]. We will try to give the review of these prospects from the point of view of the up-to-date physical and cybernetic science and its last reachings.

## Problem analysis

The principal strategic direction in 1961 has been set by lecture of Stafford Beer «On a way to the cybernetic factory». He sees a control system as some black box with a large quantity of will be organized. Depending on its internal state the black box is carried out the different functions linking its input and output. Among all these functions the optimal function exists. This function realizes its operation by optimal way according to some measure of optimality. The feedback will be organized between an output of the factory and internal state of the black box ensuring optimality of search of the internal state.

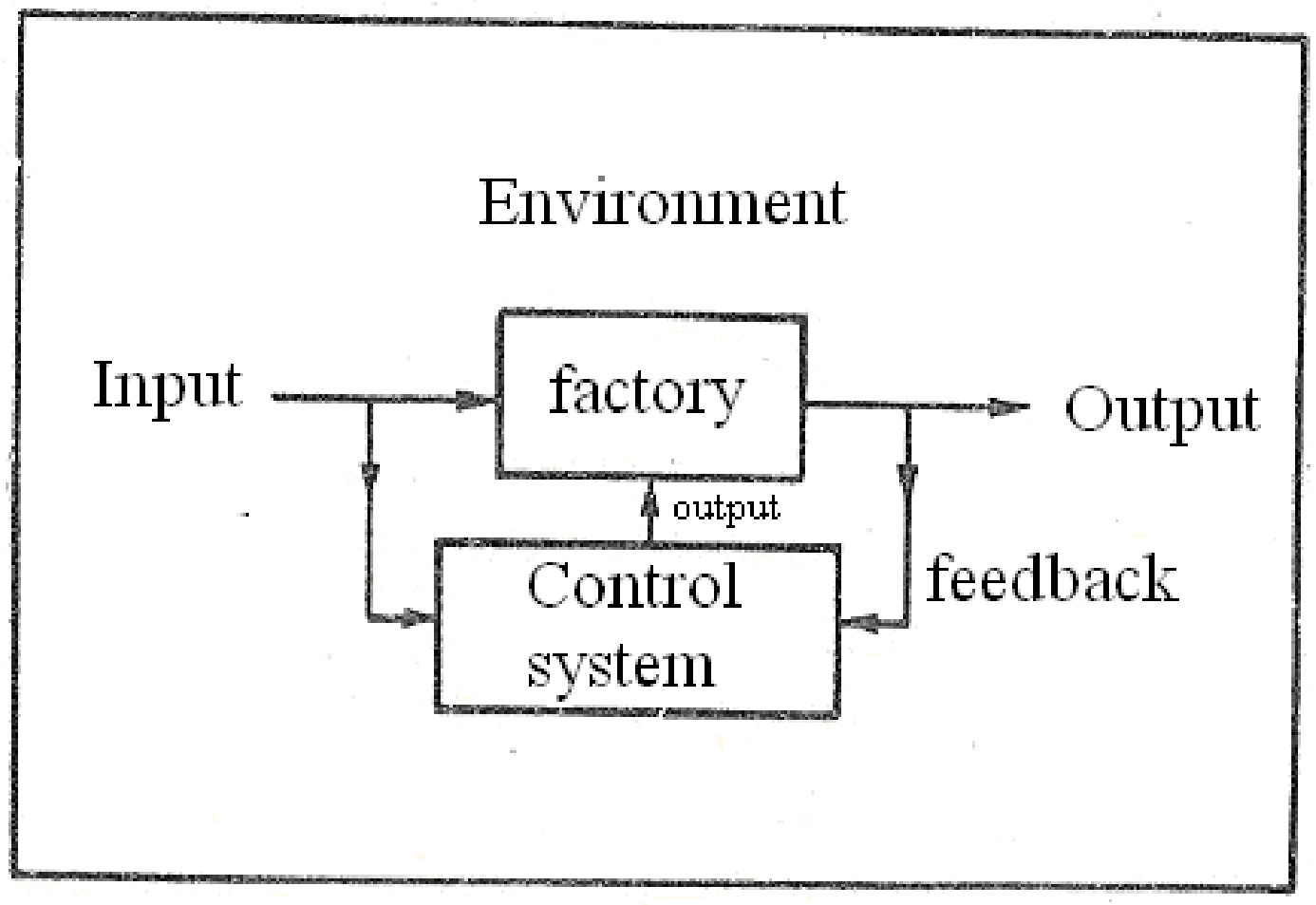

Figure 1. Diagrammatic representation of the factory controlling.

Here appear three difficulties:
1) It is clear, that the number of internal states of such black box should be huge to ensure realization of all possible functions. For this purpose the author suggests to use some block of the substance, possessing huge number of internal states at atomic level. It is something, for example, like the colloid system of Gordon Pask. This system realizes reversion of matrixes of the astronomical order.
2) Space of search of such box is huge and the search over all possible internal states is not real for reasonable time. Therefore the strategy, allowing to discover not the most optimum solutions, but, at least, "good" is necessary. Now such strategy is named as «genetic algorithm» [3], supplied *with the random generator*. Also the method of heuristics is widely used. [4] It is a set of empirical recipes for the search of optimum between the internal states. They are either found from the previous experience or defined by the external expert.
3) Criterion of optimality not always can be formulated accurately. Therefore "purpose" of such box can be made its physical "survival". Then it will search for such criterion itself. Or, its operations would be estimated by some external expert.

In the specified solutions of problems there is one very basic difficulty. Let our black box has n binary inputs and one binary output. Then number of all possible internal states of box is $2^{2^n}$ . Is how much this number great? The answer gives Willis D.G. «Set of realized functions for the complex systems». The physical calculation carried out there shows, that all molecules of the Earth is enough only for creation of the black box with maximum n=155. It does not make sense to reproduce his calculation here. The modern physics gives an exact method of calculation for the upper bound of memory through entropy of a black hole of corresponding mass [25]. (But it is problem to extract this information because of informational paradox.) The estimation for memory, however, will not be more optimistic. It is clear, that it is not enough such number of the inputs for controlling over the complex systems. Consequently the number of the possible functions, realized by box, should be some subset of all possible functions. How to choose this subset?

Now the methods based on neural networks [26] or fuzzy logic [27] actively develop. They allow to realize easily many "intuitive" algorithms which are used by people. Besides, for them there are well developed methods of training or self-training. However for both methods it is shown, that any possible function is realized by these methods. On the one hand it is good, as proves their universality. On the other hand it is bad, as this redundancy do not allow us to lower space of search of the black box, using these methods.

In his lecture Willis offers a solution, which is actual even now. He suggests to use a subset of all functions of n variables. This subset can be realized by a combination of p functions with k variables where

$$p \ll 2^n \qquad (1)$$

$$k \ll n \qquad (2)$$

This class is small enough, so it can be realized.

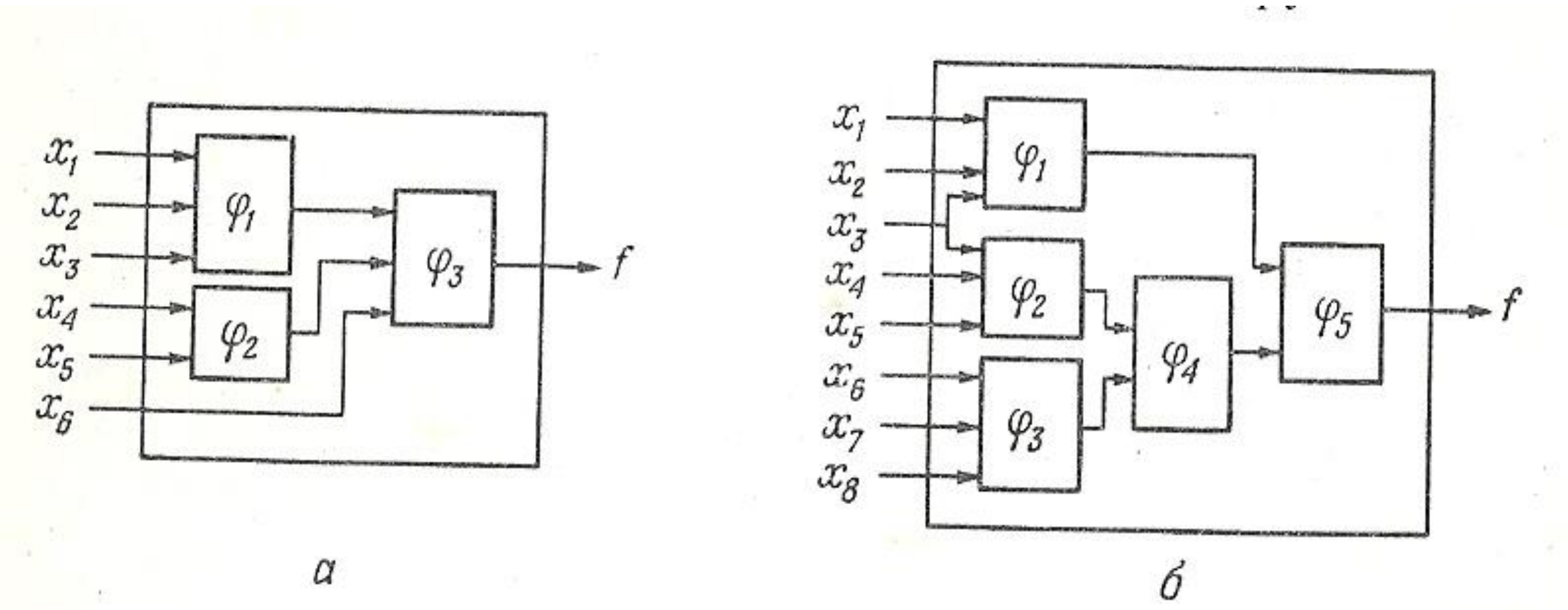

**Figure 2.** Exact expansion of switching functions on functions with smaller number of variables.
а) n=6, p=3, k=3 б) n=8, p=5, k=3

This solution is acceptable for a wide class of problems. For example, the neural network was used for recognition of the handwritten digit highlighted on the screen [28]. The screen was divided into meshes (pixels). The mesh could be black or white. Thus meshes were divided into groups of neighboring meshes (k cells). Each group arrived on input of the network with one output. These outputs were grouped also in k the nearest groups which moved on inputs of the network etc. As a result there were only 10 exits which yielded outcome of classification. The specified network uses restrictions relating to "locality" of our world.

But it is possible to introduce other similar criterions restricting space of search by less hard way. For example, we can use only the requirement (1) and not use the requirement (2). Instead of (2) we restrict type of used functions, i.e. we create some "library" of the useful functions.

For example, for existing field of the pattern recognition such set of functions already exists. It is software packages of functions for images processing. Example of such package is Matlab [29]. Combining these functions, it is possible to create a large number of the useful features for recognition. To select useful superposition of functions it is possible to use a random search of the genetic algorithm. But it can be made also using human intuition: the person can combine these functions so that they reproduced some intuitively felt feature of an object. The person himself cannot mathematically specify this feature without such search. These are human-machine systems of search.

It is worthy of note that both creation of such "libraries" and human-machine search is not algorithmizable processes. They based on human intuition. For this reason we think that the artificial intellect is closer to Art, than to Science.

Let's consider problems which appear when this approach is used:

1) Those restrictions ("libraries") which we set on internal states of the black box are human formed. It makes this process labor-consuming and restricted by human intuition.

2) Human-machine search is more effective, than the genetic algorithm, but suffers from the two above-mentioned problems.

Let's consider the following lecture which, apparently, the most prophetical and gives a trajectory to a solution of these problems: Zopf George W. «Relation and context».

His main thought is that for construction an effective model for artificial intellect we should not use some mathematical scientific abstraction like a black box. To construct such model we need use properties of similar systems in around world. These are the living adaptive systems. What their properties allow them to overcome specified above restrictions and problems?

Their most important property is that such systems are not, like a black box, some external object in relation to around world. They inseparably linked within the around world. (So, Zopf pays attention that the features used for recognition of object, or even "code" of neurons of a brain (consciousness) are context-dependent. It means that they depend not only on internal state of the object or the brain, but also their external environment.) It explains efficiency of restrictions on realized internal states of adaptive systems. They do not need to invent some "library" of search functions - it is already given them in many aspects from their birth. These systems have happened from around world and are relating to it already at their birth by a set of hidden connections. So their "library" of search functions is rather effective and optimal. The same is true for algorithms of adaptation - unlike «genetic algorithms» they are already optimally arranged with respect to around world. It allows to prevent search and verification of large number of unsuccessful variants. Moreover, "purposes" of adaptive systems are not set by somebody from the outside. They are in many aspects already arranged with respect to their search algorithms and around world restrictions.

We often perceive events in the world surrounding us as a set of independent, casual appearances. Actually, this world reminds a very complicated mechanism penetrated by a set of very complex connections. («Accidents don't happen accidentally») We cannot observe all completeness of these connections.

At first, as we are only small part of this world, it is not enough our internal states to map all its complexity. Secondly, we inevitably interact with around world and we influence him in during observation. The modern physics states, that this interaction cannot be made to naught in principle [6-12]. So to model and to consider this influence exactly we need observe not only the external world but we need observe ourselves also! Such introspection can not be made *completely in principle* at any our degree of internal complexity. Introduction of physical macrovariables only reduces acuteness of the problems, but does not resolve it.

Nevertheless, as already it was above-mentioned, we are a part of the around world and are related to it by the set of connections. So we are capable on so effective behavior. It creates illusion that we are capable effectively to foresee and to calculate everything. It is possible to name this property of adaptive living systems as superintuition[3] [13]. It considerably exceeds adaptive properties of any black box developed by purely scientific methods.

Hence, we should build our future systems of AI also on the basis of some similar "physical" adaptive systems possessing superintuition. We will give here the list of properties of such systems [9-10, 17-18].

1) The random generator of such systems (making selection of internal state) should not generate usual random numbers. Such numbers should be in the strong connection (correlation) both with the around world and with internal state of AI system, ensuring superintuition.

2) The internal state of system should be complex. It should be not equilibrium, but stationary. I.e. it should correspond to a dynamic balance. It is like a water wall in a waterfall. The internal state should be either for classical mechanics systems correlated, unstable (or even chaotic) or for quantum mechanics systems quantum coherent. Such systems are capable to conserve the complex correlations either inside of themselves or between themselves and the surround world.

3) The internal state of the system should be closed from external observation. It is reached, at first, by high internal complexity of system. Secondly, the system should change strongly the “That was Staunton’s theory, and the computer bore him out. In cases where planes or trains crash, the vehicles are running at 61 percent capacity, as regards passenger loads. In cases where they don’t, the vehicles are running at 76 per cent capacity. That’s a difference of 15 percent over a large computer run, and that sort of across-the-board deviation is significant. Staunton points out that, statistically speaking, a 3 percent deviation would be food for thought, and he’s right. It’s an anomaly the size of Texas. Staunton’s deduction was that people know which planes and trains are going to crash… that they are unconsciously predicting the future."

---

[3] The study conducted by Russian specialists under the guidance of Valeri Isakov mathematics, which specializes in paranormal phenomena. Data from domestic flights they could not be obtained, so the researchers used Western statistics. As it turned out, over the past 20 years of flight, which ended in disaster, refused on 18% more people than normal flights. "We are just mathematics, which revealed a clear statistical anomaly. But mystically-minded people may well associate it with the existence of some higher power "- quoted Isakov," Komsomolskaya Pravda ". http://mysouth.su/2011/06/scientists-have-proved-the-existence-of-guardian-angels/; http://kp.ru/daily/25707/908213/

Stephen King, "The Stand" (1990)

internal state and behavior at attempt of an external observation. This property is intrinsic for both unstable classical systems (close to chaos), and quantum coherent systems.

4) The system should be strongly protected from an external thermal noise (decoherence).

5) The system should support the classical unstable or quantum coherent state and be protected from the external thermal noise not so much passively as actively. I.e. it should not be some hard armour or low temperatures. Rather it should be some active metabolic process. The system should be in a stationary dynamic balance, instead of thermodynamic equilibrium. So the vertical wall of water in a waterfall is supported by its constant inflow from the outside.

6) The main purpose of such system should be its "survival".

To use similar systems, we need not to know in details their internal states and algorithms of operation which they will establish at interaction with around world. Moreover, trying to make it, we will strongly risk to break their normal operation. We should attend only that the purposes, which they pursue for "survival», are coincided with the solution of problems which are necessary for us.

We see that physics becomes necessary for creation of such cybernetic AI systems. Whether are there now prototypes of such systems? Many features of the abovementioned systems are inherent to the quantum computers [19-20, 24] or to their classical analogues - to the classical unstable computers [14] and to the molecular computers [16]. Besides, there is a lot of literature where the synergetic systems modeling specified above property of living systems are constructed «on a paper». In quantum field it is [21-23, 30-32], and for classical unstable systems [15].

Here appear two problems.

1) Which from above-mentioned objects will be appropriate in the best way for creation of the AI systems?

2) What purposes, necessary for "survival" of these systems, we need to put? Indeed, these purposes must be coincided with solution of our problems.

The solution of both these problems is not algorithmizable, creative process. It makes again artificial intellect to be closer to Art, than to Science. Really, usually we cannot even know how such systems are arranged inside. We can only define their restrictions only. It is necessary to direct these systems to solve problems useful for us. We often are not capable even to understand and accurately to formulate our own purposes and problems. Without all this knowledge the Science is powerless. So creation of such systems more likely will be related to writing music or drawing pictures. Only "brushes" and "canvas" will be given to us by the Science.

Whether can the AI systems solve the two abovementioned problems instead of us? For the first problem such chances exist, but the second one cannot be solved without us in principle. Indeed, none can know better than us that we want. But, both these problems are interconnected. Therefore people always will have intellectual job. It is true also for the case that our «clever assistants» will be very powerful.

## Conclusion

Perspective of an artificial intellect (AI) future is considered. It is shown, that AI development in the future will be closer to art, than a science. Complex dissipative systems which behavior cannot be understood completely in principle will be a basis of AI. Nevertheless, this not complete understanding will not be a barrier for their practical use. But a human person inevitably will conserve his important role. Completely to exclude him from the process it is impossible.

## Acknowledgment

We thank Hrvoje Nikolic and Vinko Zlatic for discussions and debates which help very much during writing this paper.

# Универсальная стрела времени V: Непредсказуемая динамика.

Купервассер О.Ю.

**Аннотация.**

Мы видим, что точные уравнения квантовой и классической механики описывают идеальную динамику, которая обратима и приводят к возвратам Пуанкаре. Реальные уравнения физики, описывающие наблюдаемую динамику, например, master equations статистической механики, уравнения гидродинамики вязкой жидкости, уравнение Больцмана в термодинамике, закон роста энтропии в изолированных системах - необратимы и исключают возвраты Пуанкаре в исходное состояние. Кроме того эти уравнения описывают системы в терминах макропараметров или функций распределения микропараметров. Причины такой разницы между динамиками две. Во-первых, неконтролируемый шум со стороны внешнего наблюдателя. Во-вторых, когда наблюдатель входит в описываемую систему (самонаблюдение) полное само-описание состояния системы невозможно. Кроме того самонаблюдение возможно в течение ограниченного времени, пока собственная термодинамическая стрела времени наблюдателя существует и не меняет своего направления. Не во всех случаях нарушенная внешним шумом (или неполная при самонаблюдении) идеальная динамика может быть заменена предсказуемой наблюдаемой динамикой. Для многих систем введение макропараметров, исчерпывающе описывающих динамику системы, просто невозможно. Их динамика становиться в принципе непредсказуемой, иногда даже вероятностно непредсказуемой. Мы назовем динамику, описывающую такую систему, *непредсказуемой динамикой*. Как следует из самого определения таких систем, для них невозможно ввести полный набор макропараметров, характерных для наблюдаемой динамики и позволяющий предсказывать их поведение. Динамика таких систем не описывается и не предсказывается *научными* методами. Таким образом, **наука сама ставит границы своей применимости.** Только сами такие системы изнутри могут, но уже *интуитивно* «понимать» и «предсказывать» свое поведение или «общаться» между собой на *интуитивном* уровне.

## 1. Введение

Дадим определения *наблюдаемой и идеальной динамик* [1-4], а также объясним необходимость введения наблюдаемой динамики. Идеальной динамикой мы будем называть точные законы квантовой или классической механики. Почему мы назвали их идеальными? Потому что для большинства реальных систем выполняется закон возрастания энтропия или редукция волнового пакета в квантовом случае, противоречащие законом идеальной динамики. Идеальная динамика обратима и в ней происходят возвраты Пуанкаре, чего не наблюдается в необратимой наблюдаемой динамике. Откуда происходит это противоречие между динамиками?

Реальный наблюдатель – это всегда макроскопическая, далекая от термодинамического равновесия система. Он обладает собственной термодинамической стрелой времени, которая существует ограниченное время (до достижения равновесия) и может менять свое направление. Кроме того, существует малое взаимодействие наблюдателя с наблюдаемой системой, которое приводит к синхронизации их термодинамических стрел времени и, в случае квантовой механики редукции волнового пакета.

Наблюдатель описывает наблюдаемую систему в терминах макропараметров и относительно собственной стрелы времени. Именно это и ведет к различию наблюдаемой и идеальной динамики, которая формулируется относительно абстрактного координатного времени в терминах микропараметров.

Нарушения идеальной динамики связаны или с незамкнутостью измеряемых систем (т.е. объясняется влиянием внешней среды или наблюдателя), или невозможностью полного само-измерения и самоанализа для замкнутых и полных физических систем, включающих как внешнюю среду, так и наблюдателя. Что же делать в таких случаях? Реальная система или незамкнута или неполна, т.е. мы не можем использовать физику для предсказания динамики системы? Отнюдь нет!

Очень многие такие системы могут быть описаны уравнениями точной (или вероятностной) динамики, несмотря на незамкнутость или неполноту описания. Мы называем её наблюдаемой динамикой. Большинство уравнений физики - master equations статистической механики, уравнения гидродинамики вязкой жидкости, уравнение Больцмана в термодинамике, закон роста энтропии в изолированных системах являются уравнениями наблюдаемой динамики.

Для того чтобы обладать указанным выше свойством наблюдаемая динамика должна отвечать определенным условиям. Она не может оперировать полным набором микропеременных. В наблюдаемой динамике мы определяем лишь много меньшее число макропеременных, которые являются некими функциями микропеременных. Это делает ее намного устойчивее по отношению к ошибкам в задании начальных условий и шуму. Действительно, изменение микросостояния не приводит неизбежно к изменению макросостояния, поскольку одному макросостоянию отвечает большой набор микросостояний. Для газа макропеременными являются, например, плотность, давление, температура и энтропия. Микропеременными же являются скорости и координаты всех его молекул.

Как из идеальной динамики строится наблюдаемая динамика? Они получаются или введением в идеальные уравнения малого, но конечного внешнего шума, или же введением погрешностей начального состояния. Погрешности и/или шумы должны быть достаточно большими, чтобы нарушить ненаблюдаемую реально обратимость движения или возвраты Пуанкаре. С другой стороны они должны быть достаточно малы, чтобы не влиять на протекание реально наблюдаемых процессов с ростом энтропии.

Для полной физической системы, включающей наблюдателя, наблюдаемую систему и окружающую среду наблюдаемая динамика не фальсифицируема в смысле Поппера [36] (при условии верности идеальной динамики). Т.е. разницу между Идеальной и Наблюдаемой Динамикой в этом случае невозможно наблюдать в эксперименте.

Однако, вполне возможны и случаи, когда введение наблюдаемой динамики невозможно и система остается все-таки непредсказуемой, вследствие незамкнутости системы или неполноты описания. Это случай *непредсказуемой динамики* [21, 29-33], обсуждаемой здесь.

## 2. Непредсказуемая динамика.

Введем понятие *синергетические модели* [10]. Будем называть таковыми простые физические или математические системы, иллюстрирующие в простой форме некие действительные или предполагаемые свойства непредсказуемых и сложных (в том числе живых) систем.

Непредсказуемые системы, именно вследствие причин своей непредсказуемости крайне неустойчивы к внешнему наблюдению и тепловому шуму. Чтобы их поведение не превратилось в полностью хаотическое и случайное, они должны иметь механизмы защиты от внешнего влияния.

Поэтому для нас важно создание в первую очередь синергетических моделей систем, способных противостоять шуму (декогеренции в квантовой механике). Они сохраняют свои внутренние корреляции (квантовые или классические), приводящие к обратимости движения или возвратам Пуанкаре. Также они могут сохранять свои корреляции с окружающим ми ром.

Существуют три метода такой защиты:

1) Пассивный метод – создание неких «стенок» или панцирей непроницаемых для шума. Можно также держать такие системы при очень низких температурах. Примером могут служить многие модели современных квантовых компьютеров.
2) Активный метод, обратный пассивному – подобно диссипативным или живым системам, они сохраняют своё неравновесное состояние благодаря активному взаимодействию и обмену энергией и веществом с окружением (метаболизмом). Думается, что будущие модели квантовых компьютеров должны браться из этой области.
3) Когда корреляции охватывают ВСЮ Вселенную. Внешний источник шума здесь просто отсутствует. Источник корреляций Вселенной в том, что Вселенная произошла из малой области и низкоэнтропийного состояния путем Большого Взрыва. Назовем это явление  глобальными корреляциями. Иногда это образно называют «голографическая модель Вселенной»

Следует отметить три обстоятельства:

1) Многие сложные системы в своем развитии проходят динамические точки бифуркации – когда существуют несколько альтернативных путей развития и выбор конкретного из них зависит от малейших изменений состояния системы в точке динамической бифуркации [5-6]. Тут даже слабые (и сохраненные, указанными выше путями) корреляции могут оказать огромное влияние. Наличие подобных корреляций ограничивает предсказательную силу науки, но отнюдь не ограничивает нашу личную интуицию. Поскольку мы являемся неотделимой частью нашего мира, то мы вполне способны на субъективном уровне «ощущать» эти корреляции, недоступные научному предсказанию (Но, ни в кое мере не противоречащие самим законам физики!)
2) В описываемых ненаблюдаемых системах часто наблюдается уменьшение энтропии или поддерживается очень низкоэнтропийное состояние. Это не противоречит второму закону возрастания энтропии. Действительно, как их пассивная, так и активная защита требуют огромных затрат негоэнтропии, которая черпается из окружения, поэтому суммарная энтропия системы и окружения только растет. Закон возрастания энтропии остается незыблемым для «большой» системы (наблюдаемая система + окружение + наблюдатель), хотя неверен для самой наблюдаемой системы. Уменьшения энтропии в большой системе согласно уравнениям идеальной динамики происходят (например, возвраты Пуанкаре в замкнутой системе с ограниченным объемом), но являются ненаблюдаемыми [1-4]. Поэтому они могут просто игнорироваться.
3) Существование многих непредсказуемых систем сопровождается уменьшением энтропии (Это не противоречит росту энтропии согласно второму закону термодинамики как это объяснено выше в третьем пункте). Таким образом, существование таких систем подчиняется обобщенному принципу Ле-Шателье — Брауна: система препятствует любому изменению своего состояния, вызванному как внешним воздействием, так и внутренними процессами, или, иными словами, — любое изменение состояния системы, вызванное как внешними, так и внутренними причинами, порождает в системе процессы, направленные на то, чтобы уменьшить это изменение. В данном случае рост энтропии порождает системы, ведущие к ее уменьшению.
4) Часто находит подтверждение принцип максимума производства энтропии (Maximum entropy production principle - MaxEPP) [38]. Согласно этому принципу неравновесная система стремиться к состоянию, при котором рост энтропии в системе будет максимальным. Несмотря на кажущееся противоречие, MaxEPP не противоречит открытому Пригожин для линейных неравновесных систем принципу минимума производства энтропии (MinEPP) [38]. Это абсолютно разные вариационные принципы, в которых хотя и ищется экстремум одной и той же функции − производства энтропии, но при этом используются различные ограничения и различные параметры варьирования. Эти принципы не нужно противопоставлять, так как они применимы к различным этапам эволюции неравновесной системы. MaxEPP означает, что диссипативные непредсказуемые системы (в том числе живые системы), находясь в замкнутых системах с ограниченным объемом, приближают наступление их термодинамического равновесия. Это значит, что они сокращают и время возврата Пуанкаре, т.е. способствую более быстрому возврату в низкоэнтропийное состояние. Это опять соответствует обобщенному принципу Ле-Шателье — Брауна: рост энтропии порождает системы, ведущие к ее уменьшению. Из всего вышеизложенного можно сделать очень интересный вывод: *глобальной «целью» диссипативных систем (в том числе и живых систем) является (а) минимизация их собственной энтропии (б) стимуляция глобальной полной системы к скорейший возврату Пуанкаре в исходное низкоэнтропийное состояние.*

5) Глобальные корреляции в общем случае «растекаются» по замкнутой системе с ограниченным объемом и ведут лишь к ненаблюдаемому возврату Пуанкаре [1-4]. Однако при наличии объектов с локальными корреляциями, глобальные корреляции могут проявляться в корреляции между такими объектами друг с другом и окружающим миром. Таким образом, наличие локальных корреляций позволяет сделать глобальные корреляции наблюдаемыми, предотвращая их полное «растекание» по системе.
6) Верное определение термодинамической макроскопической энтропии сама по себе очень трудная задача для сложных физических систем в отсутствии локального равновесия [39]
7) Следует отметить очень важное обстоятельство. Неустойчивые корреляции существуют не только в квантовой, но и в классической механике. Следовательно, подобные модели не должны носить только квантовый характер. Они могут быть и классическими! Очень часто ошибочно считается, что только квантовая механика может описывать подобные явления **[11-12].** Это не так **[7-9].** Введение «руками» малого, но конечного взаимодействия при классическом измерении и малой погрешностей начального состояния стирает разницу между свойствами квантовой и классической механики (при наличии неустойчивых корреляций микросостояний).

# 3. Синергетические модели локальных корреляций

Приведем примеры синергетических моделей непредсказуемых систем, использующих пассивный или активный метод защиты от шума.

1) Имеются исключительные случаи, для которых не происходит синхронизация стрел времени [12-13].

2) Точки фазовых переходов или точки бифуркаций. В этих точках макроскопическая система, описываемая наблюдаемой динамикой, в процессе эволюции во времени или в процессе изменения какого-либо внешнего параметра может перейти не в одно, а в несколько различных макроскопических состояний. То есть, в этих точках наблюдаемая динамика теряет свою однозначность. В этих точках возникают огромные макроскопические флюктуации, и использование макропараметров не ведет к предсказуемости системы. Эволюция становиться непредсказуемой, т.е. возникает непредсказуемая динамика.

3) Возьмем квантовую микроскопическую или мезоскопическую систему, описываемую идеальной динамикой, изолированную от декогеренции. Ее динамика зависит от неконтролируемых микроскопических **квантовых корреляций**. Эти корреляции очень неустойчивы и вследствие декогеренции (т.е. запутывания с окружением или наблюдателем) исчезают. Пусть некий первый наблюдатель фиксирует лишь начальное и конечное состояние системы. В промежутке времени между ними система полностью или почти изолирована от окружения или этого наблюдателя. В таком случае эти микроскопические корреляции не исчезают и влияют на динамику. Рассмотрим другого внешнего наблюдателя, не знающего начального состояния системы. В отличие от первого наблюдателя, знающего начального состояния системы, поведение системы для второго наблюдателя становиться непредсказуемым! Т.е. с точки зрения такого наблюдателя возникает непредсказуемая динамика. В квантовой области примерами таких систем являются *квантовые компьютеры* и *квантовые криптографические передающие системы* **[14-15].**

   Квантовые компьютеры имеют не только свойство непредсказуемости для наблюдателя, не информированного об их состоянии при запуске вычислений. Другим важным свойством является их высокая параллельность вычисления. Оно достигается за счет того, что начальное состояние является суперпозицией многих возможных начальных состояний «квантовых битов информации». За счет линейности уравнений квантовой механики эта суперпозиция сохраняется и «обработка» всех состояний, входящих в суперпозицию, происходит одновременно (параллельно). Эта параллельность приводит к тому, что многие задачи, которые обычный компьютер решает очень медленно из-за того, что рассматривает все случаи последовательно, квантовый решает очень быстро. С этим свойством и связаны надежды на практическую пользу квантовых компьютеров.

   Квантовые криптографические передающие системы используют в первую очередь свойство своей ненаблюдаемости «передаваемых квантовых сообщений» для внешнего наблюдателя, не информированного об их состоянии при начале передачи. Любая попытка прочесть передаваемое сообщение приводит к его взаимодействию с этим наблюдателем и, следовательно, «разрушению» передаваемого сообщения и невозможности прочесть это сообщение. Таким образом, перехват сообщений оказывается **в принципе невозможным** по законам физики.

4) Следует особо отметить, что, вопреки широко распространенному заблуждению, как квантовые компьютеры, так и квантовая криптография **[14-15]** имеют классические аналоги. Действительно, в классических системах в отличие от квантовых систем измерение можно провести абсолютно точно, не искажая измеряемое состояние. Однако и в классических хаотических системах имеются неконтролируемые и неустойчивые микроскопические дополнительные корреляции, обеспечивающие обратимость и возвраты Пуанкаре системы. Введём «руками» конечное, но малое взаимодействие в классическое измерение или конечную погрешность в начальные условия, которые в реальных ситуациях, и на самом деле, всегда существуют. Они стирают разницу между классической и квантовой системой. В реальных системах всегда присутствует малый внешний шум, выполняющий эту роль. Изолируя хаотическую классическую систему от этого шума, мы получаем классические аналоги изолированных квантовых устройств с квантовыми корреляциями.

   Существуют синергетические модели классических компьютеров, обеспечивающих, подобно квантовым компьютерам, невероятную параллельность вычислений [7].

   Аналогом квантовых компьютеров являются и *молекулярные компьютеры* **[9]**. Большое количество молекул обеспечивает параллельность вычислений.

Неконтролируемые и неустойчивые микроскопические дополнительные корреляции, обеспечивающие обратимость и возвраты системы, делают динамику неопределенной для наблюдателя, не информированного о состоянии компьютера в момент старта. Малое, но конечное взаимодействия при наблюдении приводит к тому, что наблюдатель нарушит нормальный запланированный ход вычислений при попытке чересчур точно померить координаты и скорости молекул, чтобы предсказать результат работы компьютера.

Аналогичные аргументы могут быть использованы для создания классических криптографических передающих систем**,** использующих явление классических неустойчивых микроскопических дополнительных корреляции. Неустранимое малое взаимодействия с перехватчиком сообщений разрушает эти корреляции. Тем самым оно делает ненаблюдаемый перехват принципиально невозможным также и в классическом случае.

5) Сохранение неустойчивых микроскопических корреляций может быть обеспечено не только за счет пассивной изоляции от внешней среды и наблюдателя, но и за счет динамического, компенсирующего помехи механизма. Это происходит в так называемых физических стационарных системах, в которых равновесие системы поддерживается за счет непрерывного потока энергии или вещества через систему. Примером могут служить микромазеры **[**16**]** – маленькие и хорошо проводящие полости с электромагнитным излучением внутри. Размер полостей настолько мал, что излучение уже необходимо описывать квантово. Оно постепенно затухает из-за взаимодействия со стенками. Эту систему оптимально описывать матрицей плотности в базисе состояний, соответствующих различным собственным энергиям системы. Этот базис наиболее устойчив к внешним шумам для любой системы близкой к термодинамическому равновесию и, следовательно, наиболее подходит для наблюдаемой динамики. Микроскопические корреляции соответствуют недиагональным элементам матрицы плотности и стремятся к нулю много быстрее, чем диагональные элементы при затухании излучения. (Иными словами, время декогеренции много меньше времени релаксации.) Однако пропускание через микромазер пучка возбужденных частиц приводит к сильному замедлению затухания недиагональных элементов матрицы плотности (иными словами микрокорреляций) и отличному от нуля стационарному излучению.

Также в теории квантовых компьютеров разработаны методы активной защиты квантовых корреляций от декогеренции, способные поддерживать их сколь угодно долго, повторяя циклы активной квантовой коррекции ошибок (QUANTUM ERROR CORRECTION). Повторение кода в квантовой информации не возможно из-за теоремы о невозможности клонирования. Peter Shor первый нашел метод квантовой корректировки ошибок, перенося информацию с одного кубита на сильно-перепутанное состояние девяти кубтов [17].

6) В физике обычно макросостояние рассматривается как некая пассивная функция микросостояния. Однако можно рассмотреть случай, когда система сама наблюдает (измеряет) как свое макросостояние, так и макросостояние окружения, записывая результат наблюдения (измерения) в микроскопическую «память». Таким путем образуется обратная связь через макросостояния на микросостояние.

Примером таких очень сложных стационарных систем являются *живые системы*. Они находятся в состоянии очень далеком от термодинамического равновесия и крайне сложны. Они упорядочены, хоть эта упорядоченность сильно отличается от периодичности неживого кристалла. Низкоэнтропийное неравновесное состояние живого поддерживается за счет роста энтропии в окружении[4]. Неравновесное состояние поддерживается за счет метаболизма – непрерывного потока вещества и энергии через живой организм. С другой стороны само это неравновесное состояние является катализатором метаболического процесса, т.е. создает и поддерживает его на необходимом уровне. Поскольку состояние живых систем является сильно неравновесным, оно может поддерживать и существующие неустойчивые корреляции, препятствуя процессу декогеренции и внешнего шума. Эти корреляции могут быть как между частями самой живой системы, так и между живой системой и другими (живыми или неживыми) системами. Если это происходит, то динамику живой системы можно отнести к непредсказуемой динамике. Несомненные успехи молекулярной биологии позволяют предсказать и описать многие черты динамики живых систем. Но нет никаких фактов, свидетельствующих, что она будет способна полностью описать всю сложность процессов в живой системе, даже с учетом ее дальнейших достижений.

[4] Так, например, растет энтропия Солнца, служащего источником энергии для жизни на Земле.

Довольно трудно проанализировать реальные живые системы в рамках концепций идеальной, наблюдаемой и непредсказуемой динамик из-за их огромной сложности. Но возможно построить гораздо менее сложны математические модели.  Это, например, неравновесные стационарные системы с метаболизмом. Это позволит понять возможную роль всех трех динамик для таких систем. Эти модели могут быть как квантовыми [11-12, 18-20, 35],  так и классическими [7-9].

7) Описанными выше случаями не описывается все многообразие непредсказуемых динамик. Нахождение точных условий, при которых идеальная динамика переходит в наблюдаемую и непредсказуемую динамику – еще полностью не решенная задача для математики и физики. Также такой еще полностью не решенной проблемой (и, по-видимому, связанной с предыдущей задачей) является роль этих трех динамик в сложных стационарных системах. Решение этих проблем позволит глубже понять физические принципы, лежащие в основе жизни.

# 4. Синергетические модели глобальных корреляций, охватывающих всю вселенную.

С помощью синергетических «игрушечных» моделей можно понять синхронистичность[5] (одновременность) причинно не связанных процессов [37], а также явление глобальных корреляций.

Глобальные корреляции Вселенной и определение живых систем, как систем, способствующих сохранению корреляций в противовес внешнему шуму, хорошо объясняет загадочное молчание КОСМОСА, т.е. отсутствие сигналов от других разумных миров. Вселенная произошла из единого центра (Большой Взрыв) и все ее части коррелированны, жизнь лишь поддерживает эти корреляции  в локальном масштабе и существует на их основе. Поэтому процессы возникновения жизни в различных частях скоррелированны и находятся на одном уровне развития, т.е. сверхцивилизаций, способных достичь Земли, пока просто нет.

## 4.1 Системы с «обострением» (blow up)

[5] Валерий Исаков, кандидат механико-математических наук и лидер небольшой группы исследователей аномальных явлений рассказал газете "Комсомольская правда" о существовании некой статистической аномалии. После того как Исакову и его группе не удалось получить данные по отказам от полетов у российских авиакомпаний, ученые воспользовались западной статистикой. Как выяснилось, за последние 20 лет от рейсов, закончившихся катастрофами, отказывались на 18% больше пассажиров, чем от благополучных.
http://kp.ru/daily/25707/908213/; http://newsru.co.il/world/23jun2011/isakov_606.html

Примером являются нестационарные системы с «обострением» (blow up) **[6,22-25],** рассмотренные школой Курдюмова. В этих процессах определяется некая функция на плоскости. Ее динамика описывается нелинейными уравнениями, подобными уравнению горения.

$$\partial\rho/\partial t = f(\rho) + \partial/\partial r(H(\rho)\partial\rho/\partial r)\,, \qquad \text{(II)}$$

где ρ - плотность, $N = \int \rho\, dr$, $r$ – пространственная координата, $t$ – временная координата, $f(\rho)$, $H(\rho)$ – нелинейные связи.

$f(\rho) \to \rho^{\beta}$, $H(\rho) \to \rho^{\sigma}$,

Эти уравнения имеют набор динамических решений, называемых решения с «обострением». Доказано существование явления локализации процессов в виде структур (при $\beta > \sigma+1$), образование дискретного их спектра с разным числом простых структур (с одиночными максимумами разной интенсивности), объединенных в несколько типов сложных структур, которые имеют различные пространственные формы и несколько максимумов. Показано, что нелинейная диссипативная среда потенциально содержит в себе спектр таких различных структур-аттракторов. Пусть ($r$, $\varphi$) – полярные координаты.

$$\rho(r,\varphi,t) = g(t)\Theta_i(\xi,\varphi), \quad \xi = \frac{r}{\psi(t)}, \; 1 < i < N$$

$$g(t) = \left(1 - \frac{t}{\tau}\right)^{-\frac{1}{\beta-1}}, \; \psi(t) = \left(1 - \frac{t}{\tau}\right)^{\frac{\beta-\sigma-1}{\beta-1}}$$

Число собственных функций:

$$N = \frac{\beta - 1}{\beta - \sigma - 1}$$

Для этих решений значение функции может стремиться к бесконечности за *конечное* время τ. Интересно, что функция достигает бесконечности в максимумах в один и тот же момент времени, то есть синхронно. По мере приближения ко времени τ решение «сжимается», максимумы «обостряются» и движутся к общему центру. В момент 0.9τ система становится неустойчивой и разрушается флюктуациями начальных условий. При высокой корреляции в начальном условии можно уменьшать эти флюктуации до сколь угодно малой величины.

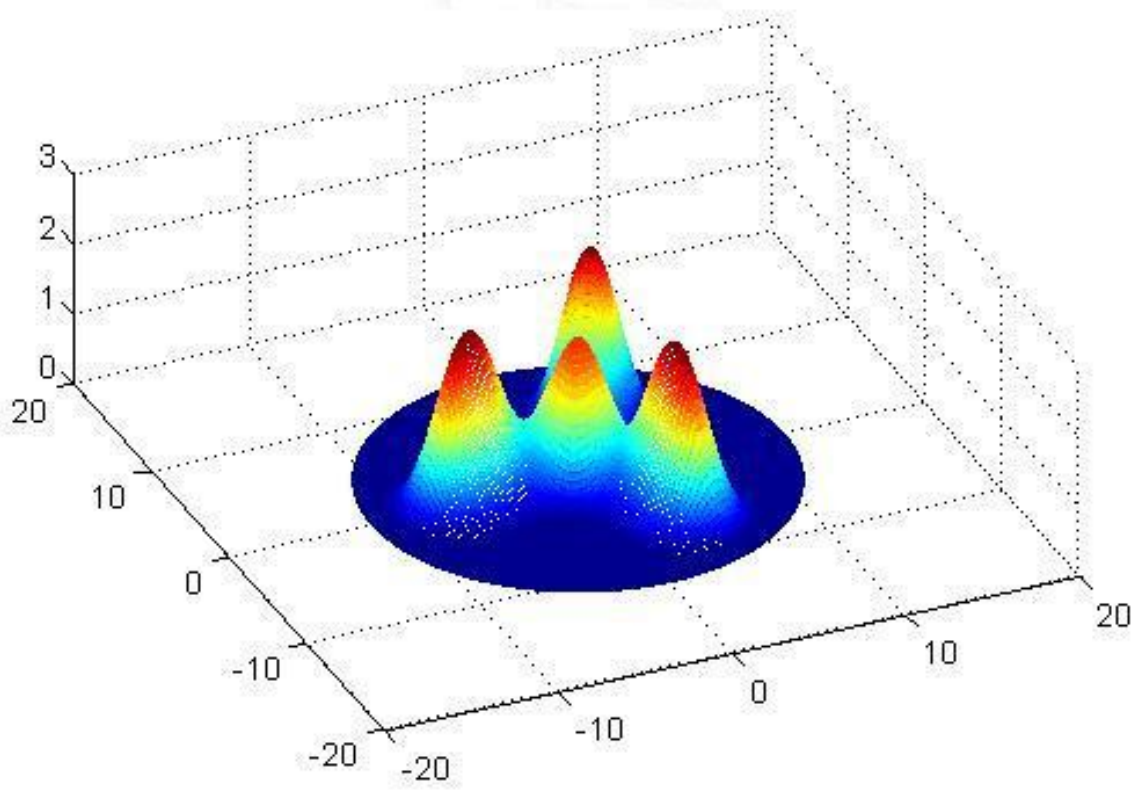

**Рис. 1** Из [34]. Один из аттракторов уравнения горения в виде решения с «обострением».

С помощью таких моделей иллюстрируют рост населения (или уровня технического развития цивилизаций) в мегаполисах нашей планеты **[25].** Точки максимума функции – это мегаполисы, а плотность населения – это значение самой функции.

Можно распространить эту модель на всю Вселенную. Тогда точки бесконечного роста – это цивилизации, а плотность населения цивилизаций (или уровня технического развития цивилизаций) – это значение самой функции. Для этого усложним модель.

Пусть в момент, когда процесс начинает выходить на растущее асимптотическое решение происходит очень быстрое расширение («инфляция») плоскости, в которой протекает процесс с «обострением». Тем не менее, процессы достижения бесконечности остаются синхронными и описываются уравнением того же типа (лишь с измененным масштабом), несмотря на то, что максимумы уже разделены большим расстоянием.

Этой более сложной моделью можно качественно объяснить синхронность развития процессов в очень далеких частях нашей резко расширившейся Вселенной в результате «инфляция» после Большого Взрыва. Высокая степень глобальных корреляций уменьшает флюктуации, ведущие к распаду структуры решения. Эти глобальные корреляции моделируют взаимосвязанность частей нашей Вселенной.

Процессы с «обострением» появляются с необходимой полнотой и сложностью лишь при некотором узком наборе коэффициентов уравнения горения ($N>>1$, $\beta>\sigma+1$, $\beta\approx\sigma+1$ – это необходимо для возникновения структуры с большим числом максимумов и их медленному сближению к центру). Это позволяет провести аналогию с «антропным принципом» **[26].** Антропный принцип утверждает, что фундаментальные постоянные Вселенной имеют именно такие значения, чтобы в итоге могла возникнуть именно наша наблюдаемая Вселенная с «антропными» существами, способными ее наблюдать.

Следует обратить внимание ещё на одно обстоятельство. Чтобы упорядоченное состояние в модели не распадалось при $t=0.9\tau$, а прожило как можно дольше, требуется тонкая настройка *не только параметров модели, но и начального состояния*. Это нужно, чтобы возникающие из него флюктуации не разрушали упорядоченность как можно дольше. И это наличие этого редкого эксклюзивного состояния также может быть объяснено антропным принципом.

## 4.2 «Клеточная» модель Вселенной.

Также интересно проиллюстрировать сложные процессы с помощью «клеточной» модели. Хорошей базой служат дискретная модель Хопфилда [27-28]. Эта модель может интерпретироваться как нейронная сеть с обратной связью или как спиновая решетка (спиновое стекло) с неодинаковыми взаимодействиями между спинами. Подобная система используется для целей распознавания образов.

Эту систему можно описать как квадратную двухмерную решетку ячеек *NxN*, которые могут быть либо черными, либо белыми ($S_i=\pm 1$). Коэффициенты линейного взаимодействия между ячейками $J_{ji}$ неравны для разных пар ячеек. Их можно выбрать так, что в процессе дискретной эволюции подавляющее большинство начальных состояний переходит в одно из возможных конечных состояний, из заранее заданного набора состояний (аттракторов).

$$S_i(t+1) = sign\left[\sum_{j=1}^{N} J_{ij} S_j(t)\right],\ 1 \le i \le N$$

$$J_{ij} = J_{ji},\ J_{kk} = 0\ \ 1 \le i, j, k \le N$$

Аттракторы соответствую минимуму энергии:

$$E = -\frac{1}{2}\sum_{i=1}^{N}\sum_{j=1}^{N} J_{ij} S_i S_j$$

Пусть аттракторами решетки выбраны буквы А или В.

Существуют такие начальные неустойчивые состояния, которые отличаются лишь на одну ячейку (*критический элемент*). При этом одно из них имеет в качестве аттрактора состояние А, а другое – состояние В. Подобные неустойчивые начальные состояния хорошо иллюстрирует свойство *глобальной неустойчивости* сложных систем. Эта неустойчивость присуща всей системе в целом, а не какой-то ее части. Лишь некий внешний наблюдатель может привести к изменению значения критического элемента и изменить эволюцию системы. Внутренняя динамика самой системы сделать это не может. *Глобальная корреляция* между ячейками неустойчивого начального состояния определяет к какому именно аттрактору эволюционирует эта решетка (либо А, либо В).

Можно несколько усложнить модель. Пусть каждая ячейка в описанной выше решетке сама является аналогичной подрешеткой. Определим эволюцию такой составной решетки, идущей в два этапа.

На первом этапе крупные ячейки не взаимодействуют, взаимодействие есть лишь в подрешетках, которое идет по тому же образцу, что и в описанной выше простой одноэтапной модели. Коэффициенты линейного взаимодействия между ячейками выбраны так, чтобы аттракторами, как и ранее, были буква А или В. Начальные состояния всех подрешеток можно выбрать неустойчивыми, содержащими критический элемент. Итоговое состояние А подрешетки будем воспринимать как черную ячейку для крупной решетки, а состояние В подрешетки будем воспринимать как белую ячейку.

Второй этап эволюции определяется как эволюция уже этой крупной решетки по тому же образцу, что и в описанной выше простой одноэтапной модели, с получившимся выше начальным состоянием. Это начальное состояние, возникающее на первом этапе, тоже является неустойчивым, содержащее критический элемент. В конце эволюции каждой черной ячейке присвоим состояние А подрешетки, каждой белой ячейке присвоим состояние В подрешетки.

Начальное состояние решетки до начало двухэтапного процесса всегда можно выбрать так, чтобы после него итоговым состоянием крупнозернистой решетки была буква А. Состоянию каждой ее крупной ячейки тоже соответствует буква А. Назовем это состояние «А-А». Тогда наличие именно такого, а не иного финального состояния можно объяснить:

a) глобальными корреляциями неустойчивого начального состояния
b) конкретным выбором всех коэффициентов взаимодействия между ячейками.

Усложним модель еще более. По аналогии с вышеописанным, сделаем ее решетку не двухуровневой, а трехуровневой, а процесс вместо двухэтапного трехэтапным. Итоговым состоянием будет «А-А-А».

Будем считать, что до начала описанного выше трехэтапного процесса, наша крупнозернистая решетка занимала очень малую область физического пространства, но в результате расширения («инфляции») расширилась до больших размеров, после чего и начался описанный выше трехэтапный процесс. Тогда наличие коррелированного неустойчивого начального состояния составной решетки, приводящего именно к итоговому состоянию «А-А-А» можно объяснить тем, что до «инфляции» все ячейки находились близко друг от друга.

Всю эту крупнозернистую решетку в целом можно сравнить с нашей «Вселенной». Ее самые мелкие подрешетки можно сравнить с «разумными организмами». Отсутствие их взаимодействия с окружением до формирования итогового состояния «А» – эквивалентно их защите от внешнего шума (активно или пассивно) своих внутренних корреляций. Решетки второго уровня в состоянии «А-А» соответствуют «цивилизациям», которые формируют образовавшиеся «разумные организмы» на втором этапе. На третьем этапе из «цивилизаций» формируется «сверхцивилизация» «А-А-А».

Тогда глобальные корреляции неустойчивых начальных состояний решеток могут служить аналогами возможных глобальных корреляций неустойчивого начального состояния нашей Вселенной, возникшего до ее инфляции. Коэффициенты взаимодействия ячеек соответствуют фундаментальным константам. Начальный процесс расширения решетки, до её трехэтапной эволюции, соответствует Большому Взрыву. Специфический выбор коэффициентов взаимодействия ячеек, приводящий к итоговой асимптотике (состоянию «А-А-А»), и начальные корреляции можно объяснять по аналогии с «антропным принципом». Антропный принцип утверждает, что фундаментальные постоянные Вселенной имеют именно такие значения, чтобы в итоге могла возникнуть именно наша наблюдаемая Вселенная с «антропными» существами, способными ее наблюдать.

## Благодарности

Я благодарю Hrvoje Nikolic и Vinko Zlatic за обсуждения и дискуссии, которые очень помогли при написании этой статьи.

## Библиография.

# Универсальная стрела времени VI:

## Будущее искусственного интеллекта - искусство, а не наука или Практическое применение непредсказуемых систем

**Купервассер О.Ю.**

Рассматриваются перспективы развития искусственного интеллекта (ИИ). Показывается, что разработка ИИ в будущем будет ближе к искусству, чем науке. Основой систем ИИ будут сложные диссипативные системы, поведение которых будет невозможно до конца понять даже в принципе. Тем не менее, это не будет препятствием для их практического использования.

### Введение

Сейчас во всем мире активно развиваются технологии, связанные с построением систем искусственного интеллекта (ИИ). В этой статье хотелось бы обсудить не тактические, а стратегические задачи этого процесса. Сейчас не так много интересных работ на эту  тему, хотя они и имеются [1]. Это связано с тем, что большинство серьезных специалистов занято решением именно тактических задач и часто не задумываются о более далеких перспективах. Однако не такова была ситуация на заре зарождения кибернетики. Тогда эти вопросы активно обсуждались. Поэтому мы построим нашу статью на обзоре задач кибернетики, как они виделись участниками симпозиума в 1961 году [2]. Мы постараемся дать обзор этих перспектив с точки зрения современной физической и  кибернетической науки и ее последних достижений.

### Анализ проблем.

Главное стратегическое направление в 1961 было задано лекцией Бира «На пути к кибернетическому предприятию».  В ней он видит систему управления как некий черный ящик с огромным количеством внутренних состояний. В зависимости от внутреннего состояния черный ящик осуществляется разные функции, связывающие его вход и выход. Среди всех этих функций ищется некая функция, оптимально реализующая его работу, согласно некоторым критериям оптимальности. Организуется обратная связь между выходом предприятия и внутренним состоянием черного ящика, обеспечивающая оптимальность поиска.

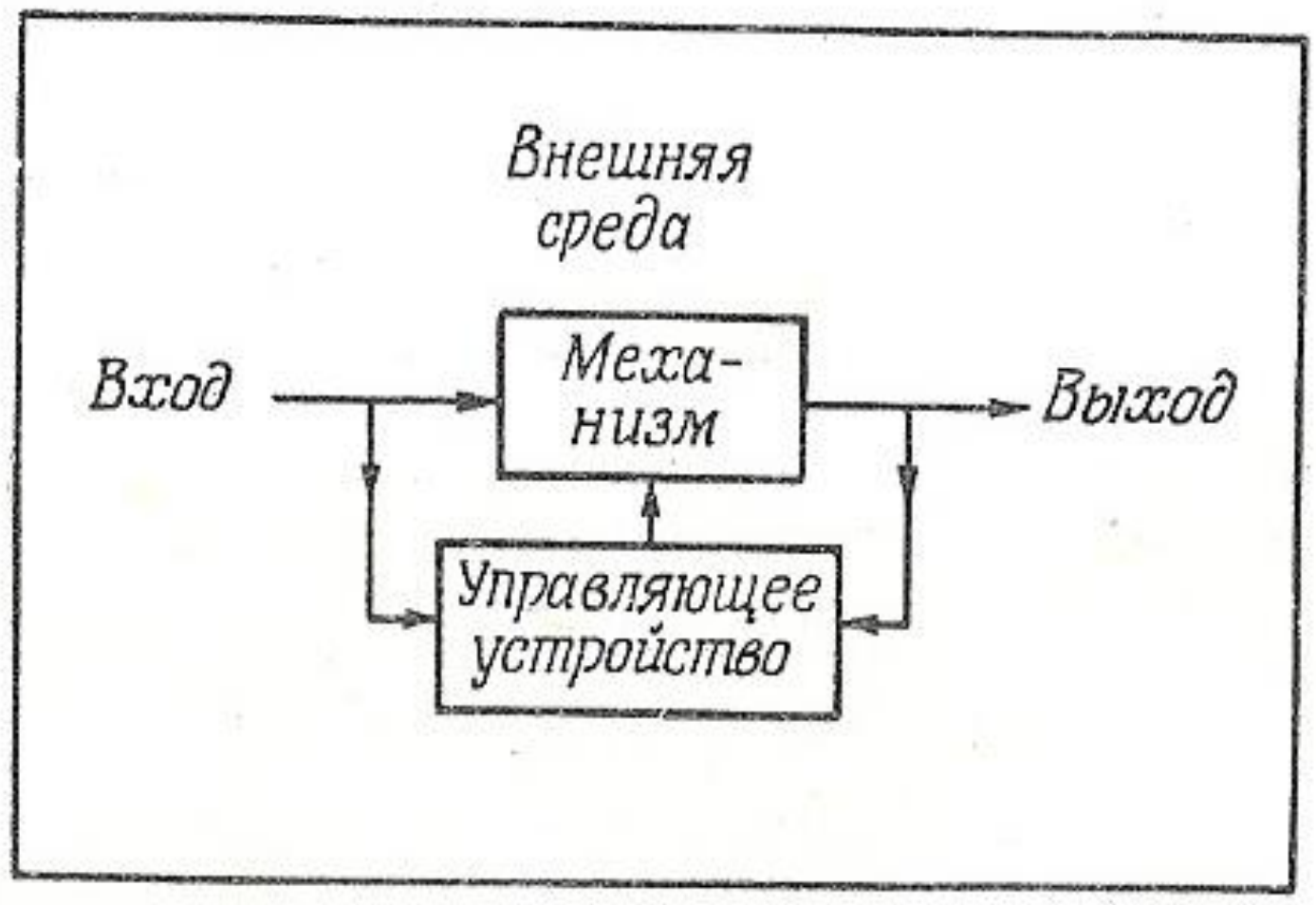

**Рис. 1** Схематическое изображение управления механизмом (предприятием).

Тут возникают три трудности:

1) Понятно, что число внутренних состояний такого черного ящика должно огромно, чтобы обеспечить реализацию всех возможных функций. Для этого автор предлагает использовать некая глыба вещества, обладающий огромным числом внутренних состояний на атомарном уровне. Это нечто, вроде, например, коллоидной системы Гордона Паска, осуществляющее обращение матриц астрономического порядка.
2) Пространство поиска такого ящика огромно и перебор всех возможных внутренних состояний за разумное время не реален. Поэтому необходима стратегия, позволяющая находить пусть не самые оптимальные решения, но, по крайней мере, «хорошие». Такой стратегией в настоящее время считается «генетический алгоритм» [3], снабженный *случайным генератором*. Также используются метод эвристик. [4] Это набор эмпирических рецептов поиска оптимального внутреннего состояния. Они либо находятся из предыдущего опыта, либо заранее заданы внешним экспертом.
3) Критерии оптимальности не всегда можно четко сформулировать. Поэтому «целью» такого ящика можно сделать просто физическое «выживание». Тогда подобные критерии он будет искать сам. Либо, его действиям будет давать оценку некий внешний эксперт.

В указанных решениях проблем есть одна очень принципиальная трудность. Пусть наш черный ящик имеет n бинарных входов и один бинарный выход. Тогда число всех возможных внутренних состояний ящика $2^{2^n}$. Насколько велико это число? Ответ дает Виллис «Область реализуемых функций для сложных систем» Проведенный им физический расчет, показывающий, что всех молекул Земли достаточно лишь для реализации черного ящика с максимум n=155 . Здесь не имеет смысл воспроизводить его расчет. Современная физика дает точный метод подсчета для верхней границы плотности хранения информации через энтропию черной дыры соответствующей массы [25]. (Ее правда проблематично извлечь из-за информационного парадокса.) Ответ, однако, вряд ли будет более утешительным. Понятно, что такого количества входов не достаточно для управления сложными системами. Отсюда следует, что количество возможных функций, реализуемым ящиком, должно быть неким подмножеством всех возможных функций. Как же выбрать это подмножество?

Сейчас активно развиваются методы, основанные на нейронных сетях [26] или нечеткой логике [27]. Они позволяют легко реализовать многие «интуитивные» алгоритмы, которые использует человек. Кроме того, для них существуют хорошо разработанные методы обучения или самообучения. Однако для обоих методов показано, что любая возможная функция реализуема этими методами. С одной стороны это хорошо, поскольку доказывает их универсальность. С другой стороны это плохо, поскольку эта избыточность не позволяют нам снизить пространство поиска черного ящика.

В своей лекции Виллис предлагает решение, которое актуально и поныне. Он предлагает использовать подмножество всех функций n переменных, которое реализуется комбинацией p функций k переменных, где

$p<<2^n$ (1)

$k<<n$ (2)

Этот класс достаточно мал, чтобы его можно было реализовать.

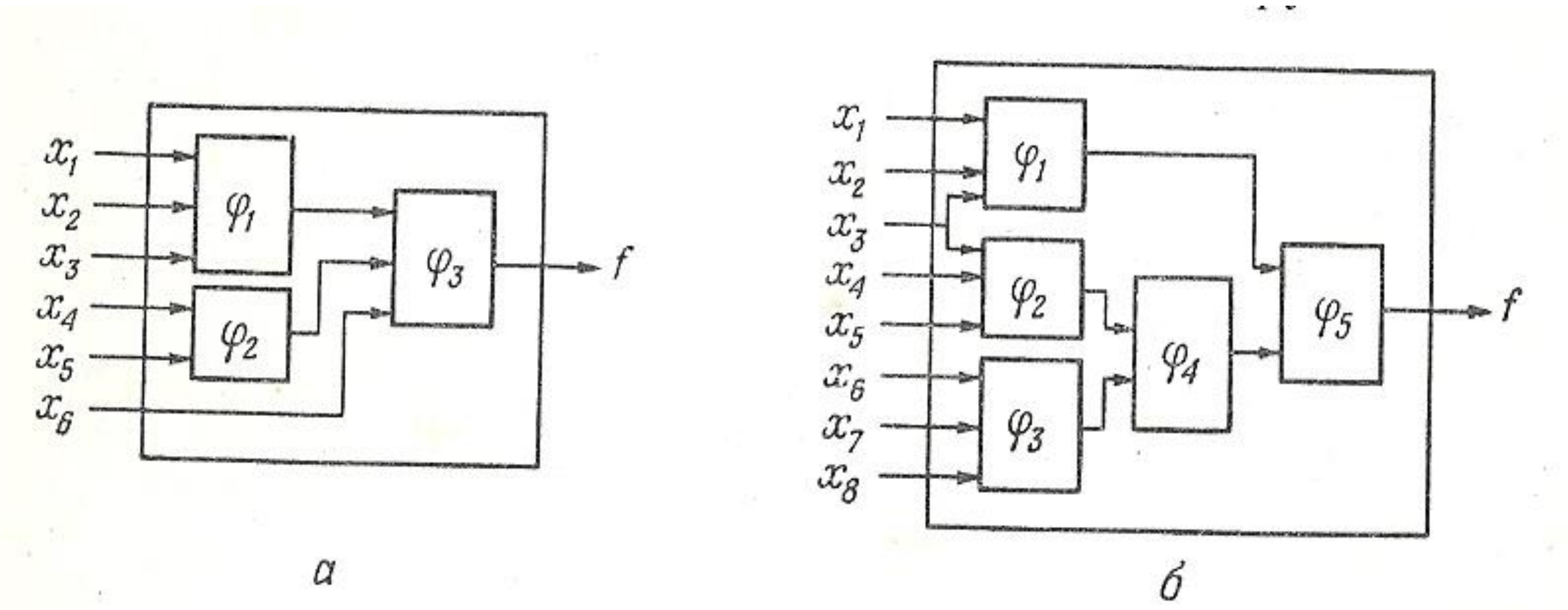

**Рис. 2** Точное разложение переключательных функций на функции с меньшим числом переменных.

а) n=6, p=3, k=3 б) n=8, p=5, k=3

Для широкого класса задач это решение приемлемо. Например, нейронная сеть использовалась для распознавания цифр, высвечиваемых на экране [28]. Экран разбивался на ячейки (пиксели). Ячейка могла быть черной или белой. При этом ячейки разбивались на группы близлежащих ячеек (k-ячеек). Каждая группа поступала на вход сети с одним выходом. Эти выходы группировались также в k ближайших групп, которые подавались на входы сетей и т.д. В итоге имелись лишь 10 выходов, которые и давали результат классификации. Указанная сеть учитывает ограничения связанные с «локальностью» нашего мира.

Но можно вводить и иные критерии ограничивающие пространство поиска того же типа, но менее жесткие. Например, сохранив условие (1), не использовать условие (2), а создать ограничение на тип используемых функций, создать некую «библиотеку» полезных функций.

Например, для существующей области распознавания изображений такой набор функций уже существует – это программные пакеты функций для обработки изображений сосредоточенные в таких пакетах, как Matlab [29]. Комбинируя эти функции, можно создать массу полезных признаков для распознавания. Причем подбирать эти суперпозиции функций можно не случайным перебором генетического алгоритма, а используя человеческую интуицию: человек может комбинировать эти функции так, чтобы они воспроизводили некий интуитивно ощущаемый признак объекта, который человек сам не может математически точно определить. Это человеко-машинные системы поиска.

Следует отметить, что как создание подобных «библиотек», так и человеко-машинные поиск – это не алгоритмизуемые процессы. Они опираются на человеческую интуицию. Именно поэтому мы считаем, что искусственный интеллект ближе к искусству, чем к науке.

Разберем проблемы, которыми страдает этот подход.

1) Те ограничения («библиотеки»), которые мы задаем на внутренние состояния ящика, создаются человеком. Это делает этот процесс трудоемким и ограниченным интуицией человека.
2) и человеко-машинный поиск, более эффективный, чем генетический алгоритм, но страдает теми же недостатками, что и описано выше.

Перейдем к следующему докладу, который, кажется, наиболее пророческий и даёт путь к решению этих проблем: Цопф «Отношение и контекст».

Главная его мысль заключается в том, что для поиска эффективных систем искусственного интеллекта мы должны обратиться не к математическим научным абстракциям, а к свойствам подобных систем в окружающем мире – живым адаптивным системам. Какие их свойства позволяют им преодолеть указанные выше ограничения и проблемы?

Самое главное их свойство заключается в том, что подобные системы не являются, как черный ящик, неким внешним объектом по отношению к окружающему миру. Они неразрывно связаны с ним. (Так, Цопф подчеркивает, что признаки, используемые для распознавания объекта, или даже сам «код» нейронов мозга (сознание) являются контекстно-зависимыми. Это значит, что они зависят не только от внутреннего состояния объекта или мозга, но также и его внешнего окружения.) Это объясняет эффективность ограничений на реализуемые внутренние состояния адаптивных систем. Им не нужно придумывать свою «библиотеку» функций – она дана им во многом от рождения. Поскольку эти системы произошли из окружающего мира и связаны с ним уже при рождении множеством незримых связей, эта «библиотека» весьма эффективна и оптимальна. То же самое относится и к алгоритмам адаптации – в отличие от «генетических алгоритмов» они уже оптимально подстроены под окружающий мир, что избавляет от перебора массы ненужных вариантов. Более того сами «цели» адаптивных систем не задаются кем-то извне. Они во многом уже подстроены под их алгоритмы поиска и ограничения окружающего мира.

Мы часто воспринимаем события в окружающем нас мире как набор независимых, случайных явлений. На самом деле, этот мир скорее напоминает сложнейший механизм, пронизанный множеством сложных связей. («Случайности не бывают случайными») Мы не можем наблюдать всю полноту этих связей. Во-первых, поскольку мы являемся лишь малой частью этого мира, то наших внутренних состояний не достаточно, чтобы отобразить всю его сложность. Во-вторых, мы неизбежно взаимодействуем с окружающим миром и влияем на него в процессе наблюдения. Современная физика утверждает, что это взаимодействие не может быть в принципе сведено к нулю [6-12]. Чтобы промоделировать и учесть это влияние нам нужно отображать внутри себя не только внешний мир, но и самих себя! Такое самонаблюдение невозможно *провести в полной мере в принципе,* при любой нашей степени внутренней сложности. Введение физических макропеременных лишь снижает остроту проблемы, но не решает ее.

Тем не менее, как уже говорилось выше, за счет того, что мы являемся частью этого мира, связаны с ним множеством связей, мы способны на столь эффективное поведение, как будто способны эффективно всё предвидеть и рассчитать. Это свойство адаптивных живых систем можно назвать сверхинтуицией[6] [13]. Оно значительно превышает адаптивные свойства любого черного ящика, разработанного чисто научными методами.

Следовательно, нам стоит строить наши будущие системы ИИ тоже на основе некоторых подобных «физических» адаптивных систем, обладающих сверхинтуицией. Дадим здесь список свойств таких систем [9-10,17-18].

1) Случайный генератор подобных систем (делающий выбор внутреннего состояния) не должен генерировать просто случайные числа. Подобные числа должны находиться в сильной связи (корреляции) как с окружающим миром, так и с внутренним состоянием системы ИИ, обеспечивая сверхинтуицию.
2) Внутренне состояние системы должно быть сложным. Оно должно быть не равновесным, а стационарным. Т.е. оно должно соответствовать динамическому равновесию, подобно водяной стене, падающего водопада. Оно должно быть или коррелированным и слабоустойчивым (даже с элементами хаоса) для систем классической механики, или когерентным квантовым для квантовой механики. Подобные системы способны поддерживать долгое время сложные корреляции между своими частями и между собой и внешним миром.
3) Внутреннее состояние система должно быть закрыто от внешнего наблюдения. Это достигается, во-первых, за счет высокой внутренней сложности системы. Во-вторых, система должна сильно менять свое внутренне состояние и поведение при попытке внешнего наблюдения. Этим свойством обладают как слабоустойчивые классические системы (близкие к хаосу), так и квантовые когерентные системы.
4) Система должна быть сильно защищена от внешнего теплового шума (декогеренции).
5) Система должна поддерживать свое классическое неустойчивое или когерентное квантовое состояние и защищаться от внешнего теплового шума не столько пассивно, сколько активно. Т.е. это не должен быть твердый панцирь или низкие температуры. Скорее это должен быть активный метаболический процесс. Система должна находиться в стационарном динамическом равновесии, а не термодинамическом равновесии. Так вертикальная стена воды в водопаде поддерживается за счет постоянного её притока извне.
6) Главной целью подобных систем должно быть их «выживание».

Для того чтобы использовать подобные системы, нам не нужно детально знать их внутреннее состояние и алгоритмы работы, которые они установят при взаимодействии с окружающим миром. Более того, пытаясь сделать это, мы будем сильно рисковать

---

[6] "Еще в 1958 году американский социолог Джеймс Стаунтон проанализировал более 200 железнодорожных аварий за предшествовавшие 30 лет. Оказалось, что поезда, закончившие свой путь трагически, в среднем были заполнены на 61% от максимально возможного числа пассажиров, тогда как в благополучные поездки отправлялись не менее 76%" Stephen King, "The Stand" (1990)

Валерий Исаков, кандидат механико-математических наук и лидер небольшой группы исследователей аномальных явлений рассказал газете "Комсомольская правда" о существовании некой статистической аномалии. После того как Исакову и его группе не удалось получить данные по отказам от полетов у российских авиакомпаний, ученые воспользовались западной статистикой. Как выяснилось, за последние 20 лет от рейсов, закончившихся катастрофами, отказывались на 18% больше пассажиров, чем от благополучных.
http://kp.ru/daily/25707/908213/; http://newsru.co.il/world/23jun2011/isakov_606.html

нарушить их нормальную работу. Мы должны лишь озаботиться, чтобы цели, которые они преследуют для своего «выживания», совпадали с нужными нам задачами.

Мы видим, что в создании таких систем физика становится необходимой для создания кибернетических систем ИИ. Имеются ли сейчас прообразы подобных систем? Многие описанные черты присущи квантовым компьютерам [19-20, 24] и их классическим аналогам – классическим неустойчивым компьютерам [14] и молекулярным компьютерам [16]. Кроме того, имеется много литературы, где «на бумаге» строятся синергетические системы, моделирующие указанные выше свойства живых систем. В квантовой области это [21-23,30-32], а для классических неустойчивых систем [15].

Тут перед нами возникает две проблемы.

1) Какие объекты, из описанных выше, будут наилучшим образом подходить для создания таких систем ИИ?
2) Какие цели, необходимые для «выживания» этих систем, нам нужно поставить, чтобы они совпадали с нашими задачами?

Решение обеих этих задач является неалгоритмизуемым творческим процессом, что опять сближает ИИ скорее с искусством, чем наукой. Действительно, зачастую мы не сможем даже знать, как устроены подобные системы внутри. Мы сможем наложить на них лишь нужные нам ограничения. Да и собственные цели и задачи мы часто не способны сами понять и четко сформулировать. Без всех этих знаний наука бессильна и создание подобных систем будет скорее сродни написанию музыки или рисованию картин. Лишь «кисточки» и «холст» даст нам наука.

Смогут ли всю эту работу выполнить за нас те же системы ИИ? Но если относительно первой задачи такие шансы есть, то вторая из этих целей вообще не может быть выполнена без нас. Ибо кто лучше нас знает, что мы хотим? Кроме того, обе эти задачи взаимосвязаны. Поэтому человеку всегда будет, чем заняться, как бы ни мощны были наши «умные помощники».

**Выводы.**

Рассмотрены перспективы развития искусственного интеллекта (ИИ). Показано, что разработка ИИ в будущем будет ближе к искусству, чем науке. Основой систем ИИ будут сложные диссипативные системы, поведение которых будет невозможно до конца понять даже в принципе. Тем не менее, это не будет препятствием для их практического использования. Но за человеком неизбежно сохраниться важная роль. Полностью исключить его из процесса невозможно

# Благодарности

Я благодарю Hrvoje Nikolic и Vinko Zlatic за обсуждения и дискуссии, которые очень помогли при написании этой статьи.

## Библиография.